\documentclass{article}
\usepackage{graphicx} 

\title{\Large Can knowledge reclassification accelerate technological innovation?}
\date{June 2025}

\usepackage[sort,comma,numbers]{natbib}
\usepackage{caption}
\usepackage{subcaption}
\usepackage{tabularx}
\usepackage{graphicx} 
\usepackage{amsmath}
\usepackage{authblk}
\usepackage[a4paper, total={6in, 8in}]{geometry}

\usepackage{nameref,hyperref}

\makeatletter
\renewcommand{\@biblabel}[1]{\quad#1.}
\makeatother

\begin{document}

\author[1, 2]{Peter Persoon}

\affil[1]{\footnotesize Oxford Martin School, University of Oxford}
\affil[2]{\footnotesize Department of Animal Sciences, Wageningen University and Research}

\maketitle

\section*{Abstract}
Technological knowledge evolves not only through the generation of new ideas, but also through the reinterpretation of existing ones. Reinterpretations lead to changes in the classification of knowledge, that is, reclassification. This study investigates how reclassified inventions can serve as renewed sources of innovation, thereby accelerating technological progress. Drawing on patent data as a proxy for technological knowledge, I discuss two empirical patterns: (i) more recent patents are more likely to get reclassified and (ii) larger technological classes acquire proportionally more reclassified patents. Using these patterns, I develop a model that explains how reclassified inventions contribute to faster innovation. The predictions of the model are supported across all major technology domains, suggesting a strong link between reclassification and the pace of technological advancement. More generally, the model connects various, seemingly unrelated knowledge quantities, providing a basis for knowledge intrinsic explanations of growth patterns.

\section*{Introduction}
Technology and science advance as new ideas are added to a cumulative structure of earlier ideas. Equally important, however, are the developments that restructure or reorganize earlier ideas, which occur for example when technologies find new applications or when scientific results are reinterpreted.\footnote{The distinction is not always clear: some relevant ideas are both new and disruptive to earlier understanding.} Think about how solar panels, once used primarily in satellites, are now installed on rooftops around the world. Or think about how Einstein's work on the photoelectric effect partly revived Newton's corpuscular theory of light. While these changes may not directly introduce new words or inventions, they potentially alter the categorization or classification of existing knowledge, i.e., they reclassify existing knowledge\footnote{Next to these intrinsic motivations to reclassify, there are extrinsic motivations to improve the classification system itself, for example, to make it more consistent, economical or searchable. However, this research is not about distinguishing between these types of motivations.} Where the societal relevance of generating new ideas is undisputed, the direct or indirect relevance of reclassification is less clear. Are reclassified ideas just passive bystanders, or are they renewed sources of future ideas? 
\begin{figure}[htbp]
\centering
\includegraphics[width=0.7\linewidth]{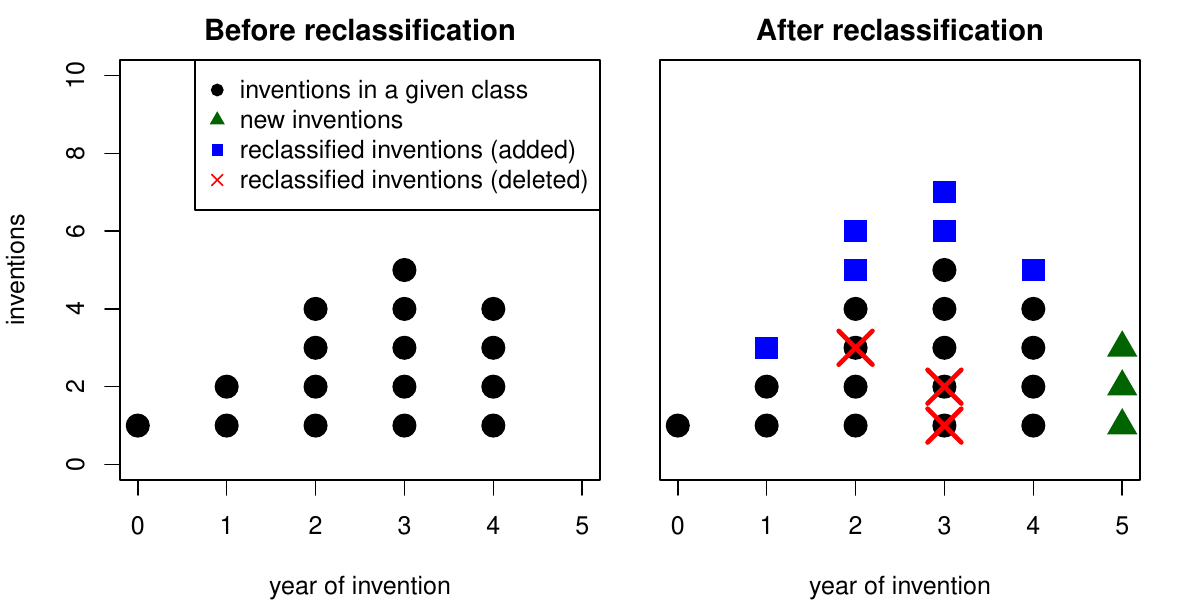}
\caption{\textit{The time development of a class of knowledge mixes two dynamics: the introduction of new inventions each year (triangles) and the inclusion/exclusion of older inventions as a result of reclassification (squares/crosses).} }
\label{schematic_growth}
\end{figure}

This is a challenging question because it inherently mixes two types of dynamics for a given class of knowledge: first, the inclusion of new ideas in a cumulative process of creation; and second, the inclusion/exclusion of older ideas as a result of ongoing reclassification; see Fig. \ref{schematic_growth}. It is likely that the two types strongly interact, but how exactly largely remains an open problem. Earlier contributions have focused on either the process of new idea generation \citep{ tria_dynamics_2014, napolitano_technology_2018, lee_idea_2024, Tacchella_predicting_2020}, or the process of classification and reclassification \citep{bailey1946history,youn_invention_2015,lafond_long-run_2019, bergeaud_classifying_2017}. This contribution demonstrates that integrating these dynamics in a single description, using patent (classification) data to proxy technological knowledge, advances our understanding of knowledge creation on various fronts.  

First, it provides an important clue as to where the large variation in growth and improvement rates across different technologies comes from \citep{Fink_how_2019, benson_quantitative_2015}. Given the close relationship between technological development, economic growth, and societal change, this continues to be one of the central issues in the innovation literature \cite{nordhaus_Economic_1969, Romer_Endogenous_1990}. Some have pointed to conceptual links between technologies as predictors of innovation rates \citep{acemoglu_innovation_2016, pichler_technological_2020}, while others have emphasized structural features such as technological complexity \citep{McNerney_Role_2011, Fink_how_2019}. Complementing this literature, this contribution establishes that technological classes that are often reclassified tend to grow faster. Although further research is needed to explore causality, the results suggest that improving classification systems may be a lever to accelerate innovation.

Second, the findings in this research have implications to the sociology and philosophy of science and technology. In theories of 'paradigm shifts', reinterpretations are the drivers of scientific \citep{kuhn_structure_1970} and technological \citep{dosi_technological_1982} revolutions. This contribution offers a quantitative, empirical perspective on reinterpretation and more generally reclassification, confirming these processes play a key part in technological developments. However, where in theories of paradigm change, reinterpretation is often presented as a short, revolutionizing event, in this contribution it is instead implemented as a routine, ongoing process.    

Third, explicitly incorporating reclassification into models of technological growth provides a fresh explanation for the commonly observed "drop" in recent patent counts—often referred to as a truncation effect \citep{hall2001nber, dass_truncation_2017}. This phenomenon is not fully explained by administrative factors, such as the 18-month lag between patent filing and publication in the U.S.\footnote{In this contribution I will not distinguish between patent applications and granted patents, the granting process can take several years.} Especially when it is focused on a particular technology such as green technology, there is a risk in interpreting this drop as a real decline in patenting rates \citep{probst_global_2021,Acemoglu_Climate_2023,aghion2022financial,martin2022private}, even though several years later the drop seems to have disappeared \citep{barbieri_evolving_2025}. This research shows that such declines can be a natural consequence of reclassification patterns and predicts the timing of these drops with high accuracy.

The paper proceeds in three parts. The first part is phenomenological, systematically observing patent reclassifications and discussing two empirical patterns\footnote{These patterns were first identified in a research project with Nicolo Barbieri and Kerstin Hotte on green technology}. The second part is analytical, where empirical patterns are used to develop a model relating reclassification to growth. The third part tests the model's predictions against a range of empirical outcomes. 

\section*{Phenomenology of reclassification} 

Patent classifications are expert-based categorizations of inventions and therefore - despite some limitations\footnote{Not all technologies are patented and many patented inventions eventually find no practical application. Also, patent classifications are primarily designed and regularly updated to assist patent examiners in their search for prior art.}- provide an ideal empirical definition of individual technologies. Furthermore, classification systems are continuously updated by adding (or deleting) classes and classifications of patents in a process called 'reclassification', thereby accommodating changing societal perspectives and technological developments that do not fit the existing scheme \citep{bailey1946history, lafond_long-run_2019}. 

\begin{figure*}[htbp]
\centering
\includegraphics[width=\linewidth]{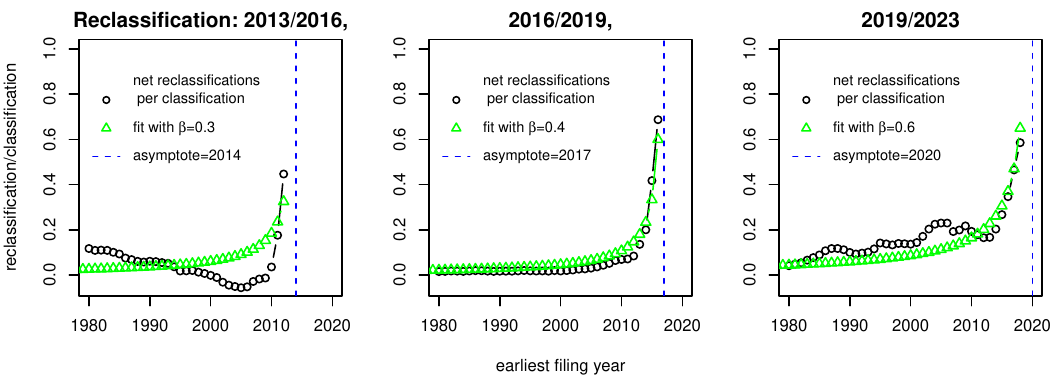}
\caption{\textit{For three reclassification moments (2013/2016, 2016/2019, and 2019/2023), I calculate the net reclassifications and divide this by the total number of classifications prior to reclassification. I plot this fraction for the earliest filing year of the corresponding patent families. As the filing years $\tau$ get close to the moment in time of reclassification $t$, the net reclassifications per classification $r(\tau,t)$ sharply increase. I include fits based on the relation $r(\tau,t)=\frac{\beta}{t-\tau}$ for different values of $\beta$. For this plot, the classifications are aggregated on subclass level. Only the families occurring in the classifications systems before and after reclassification are included.} }
\label{reclass_year}
\end{figure*}

Previous studies have pointed out several regularities in the patent (re)classifications. Research on US patent classifications showed that, since the start of the 20th century, most patents have a combination of more than one classification, that the number of distinct classes increases as a power law with the number of patents, and that the number of used combinations increases proportionally with the number of patents \citep{youn_invention_2015,lafond_long-run_2019}. Given these long-term dynamics, let me focus instead on the changes happening within a single reclassification moment. In this research, I consider patent families filed between 1970 and 2019, each including at least one US application and one filed in another jurisdiction. This comprises more than 5 million patent families, in the following also referred to simply as 'patents'. Using data from 4 Patstat editions (2013, 2016, 2019, and 2023), there are three distinct reclassification moments: 2013/2016, 2016/2019 and 2019/2023. Reclassification happens yearly, so I aggregate the changes over approximately three years for each reclassification moment. I will use the Cooperative Patent Classification (CPC) on three levels: the section level (1 digit), subclass level (four digits), and main group level (four digits plus main group number)\citep{cpc_table_nodate}. 

\begin{figure}[htbp]
\centering
\includegraphics[width=0.6\linewidth]{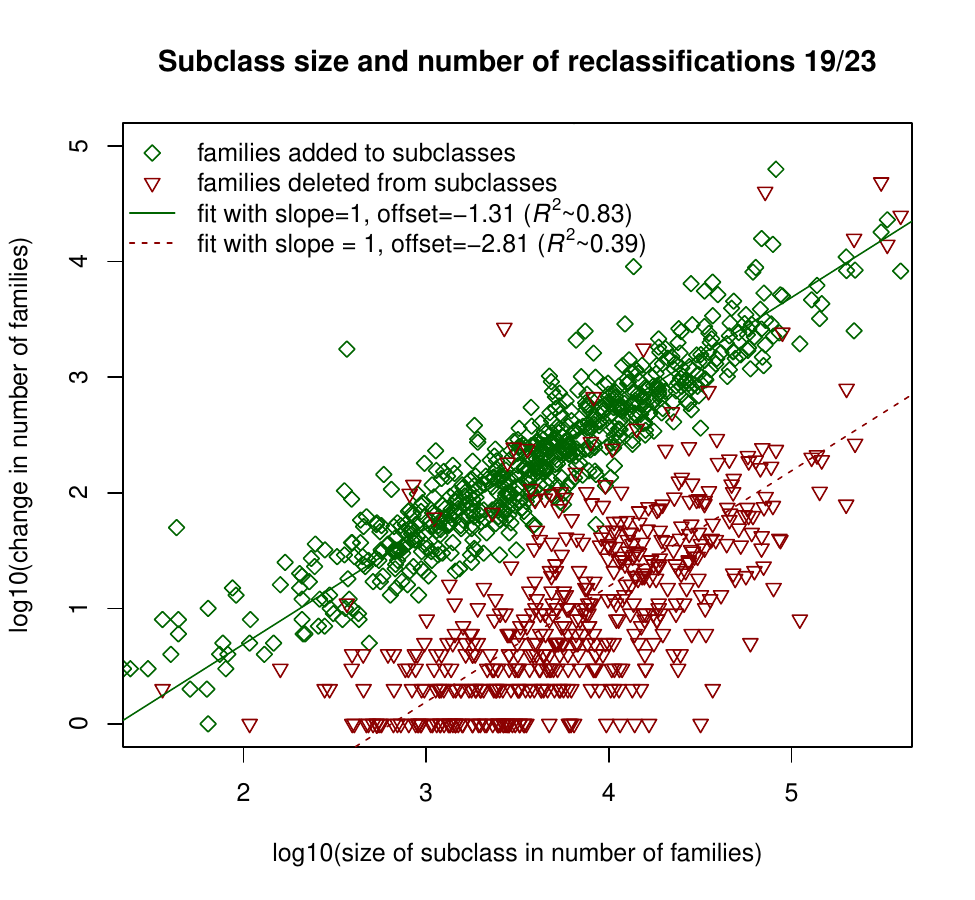}
\caption{\textit{I plot the number of families added and deleted for all subclasses for the reclassification moment 2019/2023. Note that the axes are log-transformed, hence the fits with slope $1.0$ suggest the number of reclassifications is proportional to subclass size. The offset (the y-coordinate of the fit intersection with the y-axis) that maximizes $R^{2}$ for the positive reclassifications is $-1.31$. Plotting the fit without transforming the axes thus results in a linear relation with a slope of $10^{-1.31}\approx 0.05$.} }
\label{reclass_size_classes}
\end{figure}

For a given reclassification moment and any technological class, I consider both the number of families added to that class ('positive reclassifications') and those deleted ('negative reclassifications'). The positive and negative reclassifications from each CPC subclass are plotted for the size of those subclasses (in number of families) in Fig. \ref{reclass_size_classes}. This plot focuses on the reclassification moment 2019/2023\footnote{It is almost similar for the other reclassification moments. See also the appendix \citep{barbieri_evolving_2025} for the same plot on CPC class level.}, only considering families that were already there in 2019, thus making sure that any family added is necessarily a reclassification and not a new family. This plot shows that both the positive and negative reclassifications of a class are proportional to the size of that class. For the negative reclassifications, this is intuitive: if each classification is equally likely to be deleted, the larger classes lose proportionally more families. For the positive reclassifications, however, this is more puzzling: why would larger classes attract more reclassifications? One possible explanation is that the scope of a technological class somewhat widens with each new patent, and with a wider scope, more families can be classified into that class. I leave testing this explanation for later work. 

Next, I obtain the net reclassifications for each filing year by subtracting the total negative  reclassifications from the total positive reclassifications of families with that filing year. In Figure \ref{reclass_year} I plot the net reclassifications over the total number of classifications for each filing year, each panel corresponding to a different reclassification moment. The plot shows more reclassifications per classification for patents with filing years close to the reclassification moment, in other words, recent patents are more likely to get reclassified. The pattern is similar across the three reclassification moments, except that for the 2013/2016 moment, where the net reclassifications are negative for some years. The reason for this is unclear but might relate to the fact that the CPC was still young in those years and parts were still being reorganized. A log-log plot of the positive and negative reclassifications, see the Supplementary Information (SI), indicates an inversely proportional relation between the net reclassifications per classification $r$ and the time between filing $\tau$ and reclassification $t$, i.e. $r\propto \frac{1}{t-\tau}$. Figure \ref{reclass_year} includes fits of the data based on this relation\footnote{Because each reclassification moment in the data aggregates three years, I add $r(\tau,t)= \frac{\beta}{t-\tau}$ for three subsequent values of $t$ to construct these fits, see SI appendix}, introducing the constant of proportionality $\beta$. The constant $\beta$ can be interpreted as the 'reclassification rate': a larger $\beta$ indicates more classifications are being added, while a negative $\beta$ suggests classifications are being deleted. While this relation fits the data rather well, I observe substantial variation across the three moments in $\beta$. Especially in the 2019/2023 reclassification moment, many extra classifications were added to the system. The values for $\beta$ appear to vary across different CPC (sub)classes too: some receive relatively more reclassified families than others. However, fitting and estimating $\beta$ becomes more challenging for smaller classes. In conclusion, the positive $\beta$ suggests that all of technology is increasingly more reclassified and that patents acquire more and more classifications. 

That recent patents are more likely to get reclassified makes sense for several reasons. First, when new classes are introduced, they likely consist mostly of recent patents: if they consist of mostly older patents, it would be puzzling why the class was not introduced earlier. Second, recent patents may not yet be well understood and new applications are still being discovered. Third, the language of recent patents may be more modern and the classification (search) algorithms still need some time to adjust to this language. Again, I leave it to later work to find the underlying reasons for this pattern.  

\section*{A model of reclassification and growth} 

In this section, I will introduce a model that describes the growth of a technology over time incorporating two effects: (i) how inventions in that technology trigger new inventions and (ii) how that technology is reclassified so that it includes inventions from other technologies. Let me denote the number of patents in that technology with filing year $\tau$ at time $t$ by $n_{\tau}(t)$. To incorporate (i), I build on the principle that any earlier result in a technology may at any time trigger a new result in that technology\footnote{While I do not explicitly consider the inter-technology relations in the triggering effect, note that it is implicitly accounted for in the reclassification effect} \citep{trajtenberg_university_1997,merges_limiting_1994}. The triggering of new families in a class at time $t=\tau-1$ can therefore be written as $\Delta_{t} n_{\tau}(t) \propto n(t)$, where $n(t)=\sum_{\tau}n_{\tau}(t)$.
To incorporate (ii), I explicitly use the empirical patterns identified in the previous section. The first pattern suggests that the probability for a family with filing year $\tau$ to get reclassified at time $t$ is inversely proportional to $t-\tau$. The second pattern suggests that the greater the class, the more families are reclassified into that class. The increase in $n_{\tau}(t)$ as a result of reclassification at time $t$ can thus be written as
$\Delta_{t} n_{\tau}(t)\propto n_{\tau}(t)/(t-\tau)$\footnote{I will assume that the set of different technologies or technological classes is large enough such that there can always be families reclassified to other classes.}. Finally, there is no reclassification of future families, hence $\Delta_{t} n_{\tau}(t)=0$ for $\tau-1>t$. To summarize, 
\begin{equation}\label{eqn:dynamics}
\Delta_{t} n_{\tau}(t) = 
     \begin{cases}
       0 &\quad\text{if $t<\tau-1$, }\\
       \alpha n(t) &\quad\text{if $t=\tau-1$,} \\
       \beta n_{\tau}(t)/(t-\tau+1) &\quad\text{if $t>\tau-1$,} \\
       \end{cases}
\end{equation}
where I introduce the technology-dependent constants of triggering $\alpha \geq 0$ and reclassification $\beta\geq 0$. Let me focus on the simplest case, a technology with one family at $t=\tau=0$, i.e. $n_{\tau}(0)=1$ for $\tau=0$ and $n_{\tau}(0)=0$ otherwise. As I show in more detail in the Supporting Information (SI), the third line of Equation \ref{eqn:dynamics} directly leads to the relation
\begin{equation}\label{prelim}
    n_{\tau}(t)=\binom{t-\tau+\beta}{\beta}n_{\tau}(\tau).
\end{equation}
which can be used to derive the exact solution 
\begin{equation}\label{eqn:sol1}
    n_{\tau}(t)=\binom{t-\tau+\beta}{\beta} \sum_{u=0}^{\tau}\binom{u\beta +\tau-1}{\tau-u}\alpha^{u}.
\end{equation}
Summing over $\tau\geq 0$ yields the total number of families
\begin{equation}\label{eqn:sol2}
    n(t)= \sum_{u=0}^{t}\binom{(u+1)\beta +t}{t-u}\alpha^{u}.
\end{equation}

\begin{figure}[tbhp]
\centering
\includegraphics[width=0.6\linewidth]{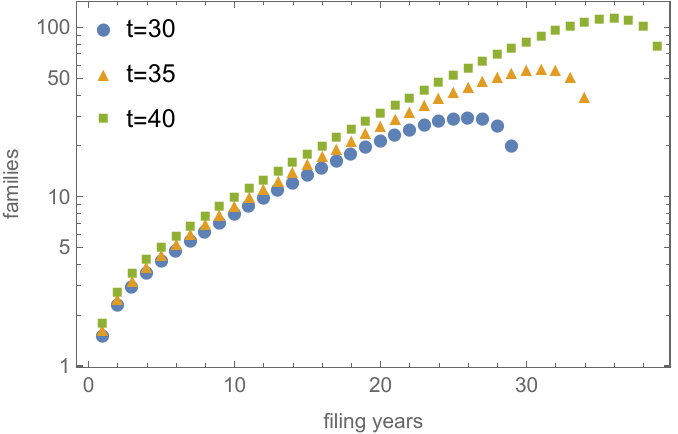}
\caption{\textit{Here is an example plot of $n_{\tau}(t)$ (see Equation \ref{eqn:sol1}) for the filing years $\tau$ where $0<\tau<t$ for $t=30,35,40$, choosing $\alpha=0.05$ and $\beta=0.5$. The number of patents increases exponentially until the apparent 'dip' as $\tau$ gets close to $t$, that is, in recent years.} }
\label{example_plot}
\end{figure}

In Figure \ref{example_plot}, I plot $n_{\tau}(t)$ for $\tau$, three values of $t$ and fixed $\alpha=0.05$ and  $\beta=0.5$. It is clear that for large $t$, the number of patents grows exponentially (note the y-axis is logarithmic) before their numbers drop again in years close to $t$. To better understand how the base of the exponential relation depends on $\alpha$ and $\beta$, it is instructive to calculate the generating function $G(z)$ of $n(t)$, that is, $G(z)=\sum_{t=0}^{\infty}n(t)z^{t}$ for some $0\leq z<1$. In the SI appendix, I demonstrate that this evaluates to
\begin{equation}
    G(z)=\frac{1}{(1-z)^{1+\beta}-\alpha z}.
\end{equation}
Furthermore, following a theorem from generating functions theory (\citep{graham_concrete_1993}, see page 341), I show that this rational generating function allows for an asymptotic estimate of $n(t)$. For real $\beta>0$ and $0<\alpha<1$, it counts
\begin{equation}\label{approx_1}
n(t) \simeq n_{0}g^t,
\end{equation}
where $n_{0}$ is a constant depending on $\alpha$, $\beta$, and the initial conditions and $g$ is a real number satisfying
\begin{equation}\label{eqn:basis}
    \left(1-\frac{1}{g}\right)^{1+\beta}= \frac{\alpha}{g}.
\end{equation}
While this implies one cannot obtain a closed expression of $g$ in terms of $\alpha$ and $\beta$, Equation \ref{eqn:basis} is enough to get a qualitative understanding of the effect of reclassification on growth. As I show in the SI, $g$ is always larger for larger $\alpha$ and larger $\beta$ and $1+\alpha\leq g < 1+\alpha +\beta$. When both $g-1<<1$ and $\alpha<<1$, i.e. for slow growth, $g\approx 1+\alpha^{\frac{1}{1+\beta}}$. This suggests that, when growth is slow, a linear increase in reclassification (that is, in $\beta$) leads to a stronger than linear acceleration of growth. Note that the upper bound $g < 1+\alpha +\beta$ implies that for larger growth factors and $\alpha$, there is at most linearity in $\beta$ instead.  

Using the model solutions, I can calculate several interesting quantities that allow for empirical validation of the model in the next section. Let me first consider the year for which the number of patents peaks before they (seemingly) decline in recent years, again see Fig \ref{example_plot} for an illustration. As is clear from this figure, the peaks are not stationary but instead shift to more recent years as $t$ increases: the decline is therefore only apparent. Let me study the typical 'decline-time' $T=t-\tau$ between filing year $\tau$ where the peak is and the present $t$. If the peak occurs at filing year $\tau$, then that is the earliest year such that $n_{\tau}(t)\geq n_{\tau+1}(t)$, or, using the relation \ref{prelim},
 \begin{align}\label{Tspan2}
    \binom{t-\tau+\beta}{\beta} n_{\tau}(\tau)&\geq\binom{t-\tau+\beta-1}{\beta} n_{\tau+1}(\tau+1).
\end{align}
From the initial conditions and the second line of Equation \ref{eqn:dynamics}, note that $n_{\tau+1}(\tau+1)=\alpha n(\tau)$. Approximating $n_{\tau+1}(\tau+1)=\alpha n_{0}g^\tau$, substituting in Equation  \ref{Tspan2} and simplifying then gives  
\begin{align}
     \frac{\beta}{t-\tau}&\geq g-1
\end{align}
The $\tau$ for which the left- and right-hand sides are equal is the peak year, hence the typical time-span between peak and present can be approximated by
\begin{equation}\label{Tspan}
    T\approx \frac{\beta}{g-1}.
\end{equation} 

Second, I consider the total number of reclassified families in year $t$, denoted by $v(t)$. The phenomenology in the previous section requires that $v(t)$ is proportional to the total number of families. Let me show the proportionality is indeed predicted by the model and at the same time derive an explicit expression for the constant of proportionality $V$, which I will refer to as the 'reclassification proportion'. Equation \ref{eqn:dynamics} leads to $v(t)=\sum_{\tau=0}^{t-1}\beta\frac{n_{\tau}(t)}{t-\tau}$. Rather than doing the summation, let me calculate the total number of families added in year $t$ by summing Equation \ref{eqn:dynamics} over all $\tau$, obtaining
\begin{align}
    \sum_{\tau=0}^{t}\Delta_{t}n_{\tau}(t)&=\alpha n(t)+\beta\sum_{\tau=0}^{t-1}\frac{n_{\tau}(t)}{t-\tau}\\
     \Delta_{t}n(t)&=\alpha n(t)+v(t).
\end{align}
Dividing left and right by $n(t)$, using the approximation for $n(t)$ in Equation \ref{approx_1} and summing over $t$, this becomes
\begin{align}
    (g-1)t&\simeq \alpha t+\sum_{t'=1}^{t} \frac{v(t')}{n(t')}.
\end{align}
For the left- and right-hand side to agree for large $t$, I conclude that $v(t)\simeq n(t)\left(g-1-\alpha\right)$. As required, the number of reclassified families is proportional to the total number of families; furthermore, the proportion of reclassification is expected to be $V= g-1-\alpha$.  

Finally, the current formulation of the model is focused on the development of one individual technological class, where $n_{\tau}(t)$ represents a set of unique patents with filing year $\tau$ at time $t$. Note that, if this set would be the collection of all patents, i.e. 'all technology'\footnote{To be precise, 'all patented technology'}, then $\beta=0$ because there are no external classes or patents to draw from. However, approaching this slightly differently, the model can still be sensibly applied to the union of all technologies: if $n_{\tau}(t)$ instead represents all unique classifications of patents with filing year $\tau$ at time $t$ (that is, counting patents with multiple classifications multiple times), then $n_{\tau}(t)$ does increase for $\tau<t$. In this interpretation, there is a relevant quantity $W_{0}(t)$ describing on average how many classifications each new patent has (upon introduction). Assuming $W_{0}$ is constant (or changes very slowly), the number of unique patents is then given by $\sum_{\tau}n_{\tau}(\tau)/W_{0}$, allowing us to calculate the number of classifications per family, denoted by $W$ (for more details see SI appendix),
\begin{align}\label{eqn:rho}
    W&=\frac{n(t)}{\sum_{\tau=0}^{t}n_{\tau}(\tau)/W_{0}},\\
          &\approx W_{0}\frac{g-1}{\alpha}.
\end{align}
Another interpretation of this expression is that the growth rate $g-1$ is proportional to $W$ (the proportional constant being $\frac{\alpha}{W_{0}}$): in other words, technologies or technological classes with more classifications per family are expected to grow proportionally faster. 

\section*{Model validations} 

In this section, I explicitly check the main predictions made by the model using patent classification data. More specifically, based on the estimated values for the model parameters $\alpha$ and $\beta$, we will compare the predicted values with the empirical values for four interesting quantities. 
\begin{enumerate}
    \item the overall growth factor $g$, 
    \item the decline-time $T$,
    \item the reclassification proportion $V$,
    \item the number of classifications per patent $W$. 
\end{enumerate} 
Let me apply the model to 'all technology' (see the last part of the previous section). Based on the definitions in Equation \ref{eqn:dynamics}, patent numbers between 2010-2015, and the fits in Fig. \ref{reclass_year}, the values for $\alpha$ and $\beta$ can be directly estimated (for a detailed explanation, see the SI appendix). I find $0.024<\alpha < 0.027$ and $\beta\approx 0.4$.   

\begin{figure}[htbp]
\centering
\includegraphics[width=0.7\linewidth]{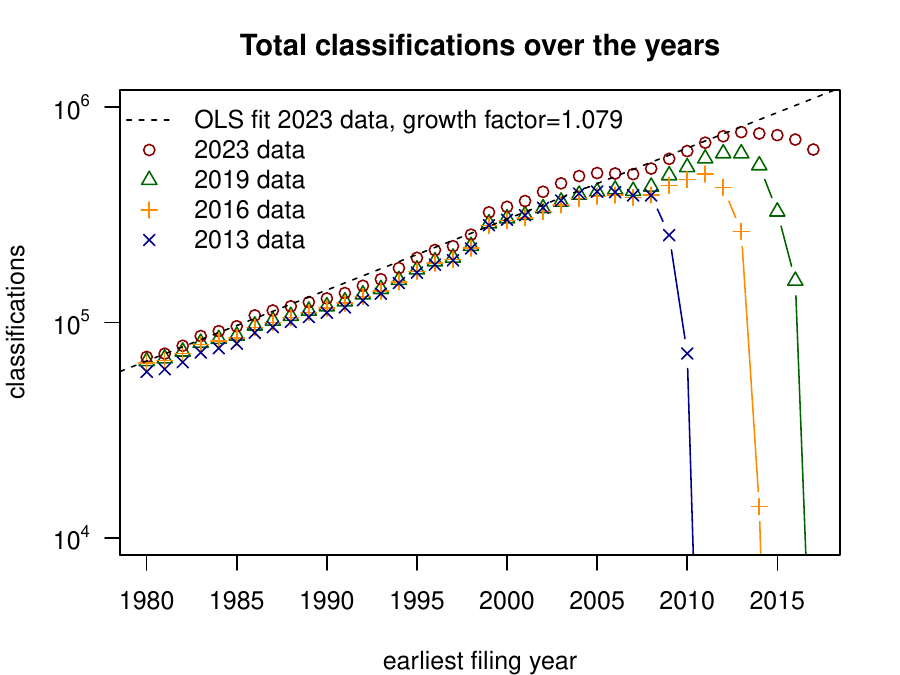}
\caption{\textit{I plot the total number of classifications of patents by earliest filing year according to 4 different datasets. I include a fit of the 2023 data ($R^2=0.98$), which indicates that classifications grow exponentially (the y-axis is logarithmic). Note the similarity with Figure \ref{example_plot}, especially with the apparent declines in recent years. A closer examination points out that the 2013, 2016, 2019, and 2023 data peak respectively in 2006, 2012, 2013, and 2014.} }
\label{decline}
\end{figure}

As a first validation, using $0.024<\alpha<0.027$ and $\beta=0.4$ in Equation \ref{eqn:basis}, I predict an overall growth-factor of $1.07 < g < 1.08$. Measuring $g$ using OLS fits for the log number of classifications between 1980 and 2015 using the 2023 dataset gives $g\approx 1.079$, see Figure \ref{decline}. For the other datasets and using the cumulative number of classifications instead, I find similar values for $g$ between $1.075$ and $1.08$. All of these are very well fitted by an exponential relation ($R^2\approx 0.98$). I conclude that Equation \ref{eqn:basis} agrees rather well with observation. 

As a second validation, I use $g=1.079$ and $\beta=0.4$ in Equation \ref{Tspan} to predict the decline-time $T\approx 5.1$. Figure \ref{decline} plots the total number of classifications of all patents by their earliest filing year and clearly demonstrates the apparent declines in recent years. Note also the similarity with Fig. \ref{example_plot}. A closer examination points out that classifications in the 2013, 2016, 2019, and 2023 data peak respectively in 2006, 2012, 2013, and 2014. This leads to decline-times in these datasets of 7,4,6 and 9 years. These values are close to the predicted value for $T$, the only exception being 2023. However, if I choose $\beta=0.6$ for the 2023 data (see Fig. \ref{reclass_year}), the prediction is that $T\approx 7.6$. 

As a third validation, I use the estimated (lower) values for $g$ and $\alpha$ in the expression $V=g-1-\alpha$ to predict a higher estimate for the reclassification proportion $V\approx 0.056$. The easiest way to measure this value is to determine, for a given reclassification moment, the total net reclassifications and divide by the number of classifications before reclassification. Using reclassification moments 2016/2019 and 2019/2023, this leads to a reclassification proportions 0.023 and 0.053 respectively. Those values are reasonably close to the predicted higher estimate for $V$. Furthermore, as explained in Figure \ref{reclass_size_classes}, the slope of the linear relation between the subclass reclassifications and size is of the order 0.05. While this value takes into account three reclassification years, the order of magnitude is in agreement with the predicted $V$. This suggests that the model applies reasonably well also to individual subclasses, that is, to individual technologies.  

\begin{figure*}[tbhp]
\centering
\includegraphics[width=\linewidth]{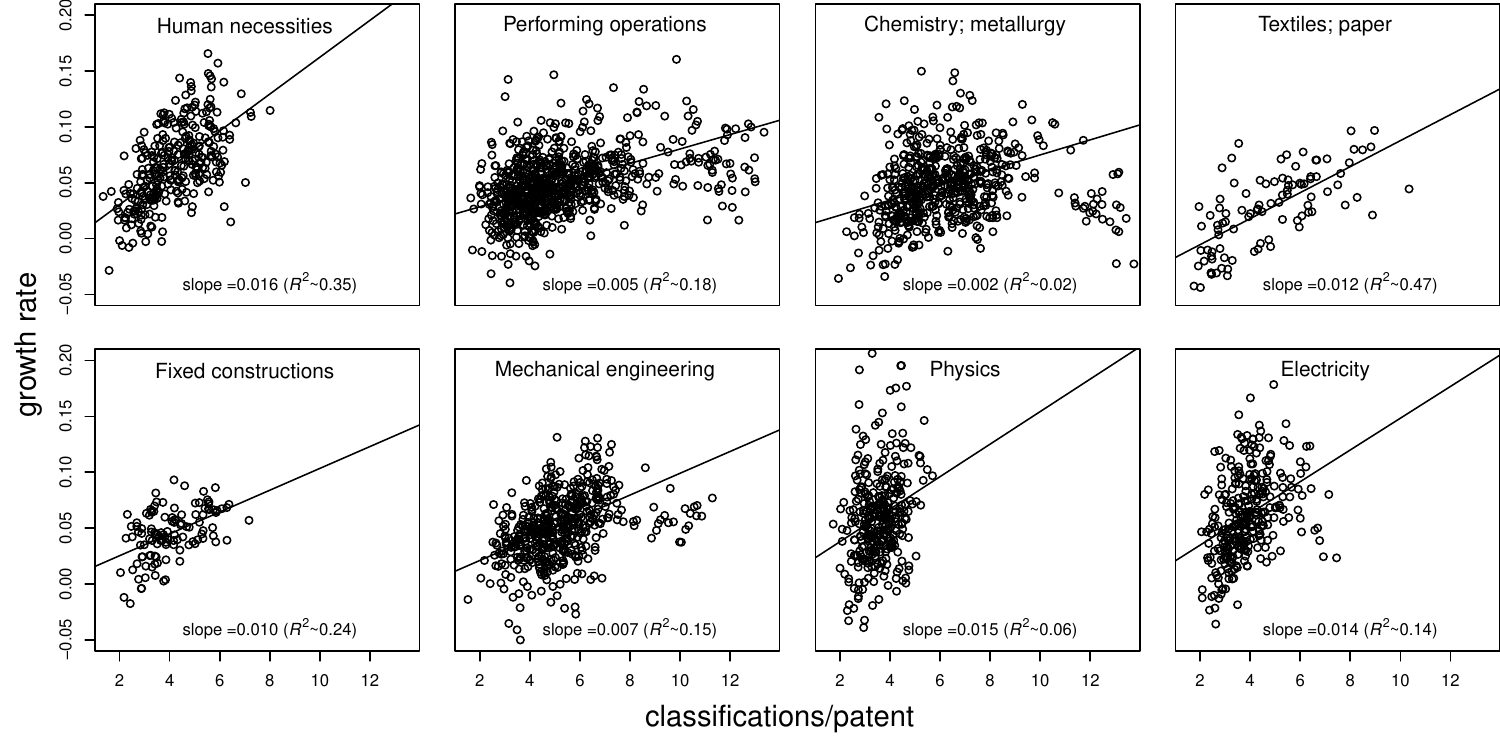}
\caption{\textit{For each CPC group $k$, I plot the growth rate $g_{k}-1$ for the average classifications per patent $w_{k}$. I plot the groups for each CPC section separately (each group belongs to one section). As predicted, most slopes are of the same order of magnitude as the estimated $\alpha\approx 0.023$.} }
\label{sections_growth}
\end{figure*}

For the fourth and final validation, I will consider the classifications per patent first for all technology and then specifically for different technologies. To predict this quantity for all technology, I will use Equation \ref{eqn:rho} and the measured values for $g-1\approx 0.079$, $0.024<\alpha<0.027$, $W_{0}=1.25$ for the CPC group level (see SI appendix) to derive that $3.7<W< 4.1$ classification/patent. Using $W_{0}=0.7$ for the CPC subclass level, this is $2.0<W< 2.13$ instead. Measuring the average number of classifications per patent (using the 2023 data) gives 3.6 classifications per patent on the CPC group level and gives 2.14 classification/patent on the CPC subclass level. The measured values are therefore very close to the predicted intervals. In the following, I will explicitly calculate the number of group classifications per patent $w_{k}$ for each CPC group $k$ (considering classifications on the CPC group level). Equation \ref{eqn:rho} predicts that this is proportional to the growth rate $g_{k}-1$ of these groups. Testing if variation across the groups in $w_{k}$ correlates with variation in $g_{k}$ is therefore an indirect test of the effect of reclassification. I sub-select all groups with at least one patent each year between 1970 and 2015 and determine the growth factor $g_{k}$ using an OLS estimation on the log-transformed number of patents each year. The number of classifications per patent $w_{k}$ is determined by dividing the total number of group classifications of each patent in a group by the number of unique patents in that group. Each group belongs to exactly one CPC section. In Figure \ref{sections_growth}, I plot each group's $w_{k}$ and $g_{k}-1$ separately for each section. For each section, I observe a highly significant and moderately strong correlations between these quantities (see the values for the Pearson correlation coefficient $R^2$). The only exception is Chemistry/metallurgy, however, as I show in more detail in the SI appendix, the correlations become much stronger when groups with $w_{k}>9$ classification/patent (i.e. very high $w_{k}$) are excluded. Those groups belong to 4 subclasses only, a very small minority. In any case, the relation between $g_{k}-1$ and $w_{k}$ appears to be especially strong for $w_{k}<9$. The relations are still highly significant if I control for the total number of patents in each group and if I control for the number of patents in recent years. This indicates that the correlations are not an indirect result of group size or having many recent patents. In the SI appendix, I include the details of two more robustness checks. In the first, I reformulate $w_{k}$ to account for the bias of recent years in reclassifications. In the second, I reformulate $g_{k}$ to exclude double counting of classifications using a fractional attribution. Even with these reformulations, I observe a strong relation between classifications and growth. Finally, it is interesting to note that the slopes of the observed linear relations, which the model predicts correspond to $\alpha/W_{0}$, appear to vary across different sections: the slopes are large for Human Necessities, Physics, and Electricity and small for Performing Operations and Chemistry. It is not directly clear where these technology-dependent variations come from. In line with prediction, note that the slopes are of the same order of magnitude as the measured $\alpha$. 

\section*{Discussion}

This research develops a model relating technological reclassification to the rate of development. Using patent classification data, I find a positive relationship between the degree of reclassification a technology undergoes and its rate of growth, consistent with the model's predictions. This suggests that further improvements to categorization and classification could significantly speed up technological innovation, and that reclassified ideas can be considered renewed sources of future ideas. There are, however, also some limitations to this research that I should mention. 

First, it is important to stress that the observed positive relationship is no direct evidence of causality. To explore this further, it is essential to better understand the mechanism proposed in the model, where earlier patents in a given class 'trigger' the development of later patents within the same class. Reclassification could then accelerate the process, as reclassified patents would trigger new results in multiple classes. Although the 'triggering event' is difficult to observe, earlier work finds a positive relationship between reclassification and increased citation rates \citep{lafond_long-run_2019}, suggesting that reclassified patents may indeed have a greater triggering potential. However, the extent to which (patent) citations can be interpreted as knowledge flows is still a matter of debate \citep{alcacer_patent_2006,SHARMA201731}. 

Second, one should keep in mind the various facets of reclassification: some reclassifications reflect deep conceptual shifts and reinterpretations, while others merely facilitate searching within the patent system and involve less substantial conceptual change. Although the distinction is relevant to the implications of this research, it is difficult to separate these facets empirically. 

Third, I should mention that, although the CPC classification is by now widely adopted, there are also other patent classifications systems in use across various countries, most notably the International Patent Classification (IPC). While validating these results with the IPC would be valuable, the CPC is largely derived from the IPC and offers more detailed classifications in many respects.   

Finally, I explore the broader implications of this research. This research indicates that reinterpretation, here considered a part of reclassification, is closely linked to the speed of development. This finding aligns well with theories of paradigm change (e.g., \citep{kuhn_structure_1970, dosi_technological_1982}), which propose that innovation accelerates during paradigm shifts, with reinterpretation playing a central role. In contrast, this study treats reclassification as a continuous process, occurring at a relatively constant rate throughout a technology's evolution and as such allowing for verifiable predictions. This raises the question: should reinterpretation be viewed as a singular event or as a continuous, everyday process? 

The results also have some direct implications for research on technological change using patent classifications. The proposed model, based on only two parameters, can explain various seemingly unrelated quantities, such as growth rates, decline-times, and the number of classifications per patent. It is intuitive to think that these quantities are determined mainly by external factors, such as fluctuations in investments or resource price dynamics. For example the earlier mentioned green decline, which ran over several years, has been attributed to factors such as falling fossil fuel prices and the financial crisis (for an overview see \citep{barbieri_evolving_2025}). Yet, as reclassification (i.e. $\beta$) for green technology is relatively strong \citep{barbieri_evolving_2025}, Equation \ref{Tspan} directly suggests that the decline-time for green technology is relatively large, offering an alternative explanation for the years of green decline. This demonstrates that it is not always necessary to bring in external factors; instead knowledge intrinsic factors might provide a simpler explanation.  

\section*{Acknowledgments}
The author thanks François Lafond for useful comments on the script. This project arose as a spin-off of a wider research project on patent reclassification together with Nicolo Barbieri and Kerstin Hotte. The author thanks Nicolo and Kerstin for joint developments of the data strategy and fruitful discussions.

%
%
%


\bibliography{reinterpretation_bib}

\section*{Supporting information}

In this Supporting Information, I first discuss how to derive the model solutions, an expression for the growth factor, and the number of classifications per family. Second, I discuss the estimation of the model parameters $\alpha$ and $\beta$ based on data. Third, I include a log log plot with the empirical analysis. Fourth, I discuss in detail three robustness checks as part of the model validation.  

\subsection*{Model solution}
This section provides a detailed derivation of the model solutions presented in the main text. Let us quickly recap on the terminology. For some technological class, let us denote the number of families with filing year $\tau$ at time $t$ by $n_{\tau}(t)$ and  $n(t)=\sum_{\tau}n_{\tau}(t)$. For convenience, let $t =0,1,2,3,..$ and $\tau =0,1,2,3,..$. The solution should satisfy
\begin{equation}\label{eqn:dynamics_a}
\Delta_{t} n_{\tau}(t) = 
     \begin{cases}
       0 &\quad\text{if $t<\tau-1$, }\\
       \alpha n(t) &\quad\text{if $t=\tau-1$,} \\
       \beta n_{\tau}(t)/(t-\tau+1) &\quad\text{if $t>\tau-1$,} \\
       \end{cases}
\end{equation}
with constants $\alpha \geq 0$ and $\beta\geq 0$. We focus on the simplest case, a technology with one family at $t=0$ and $\tau=0$, i.e. $n_{\tau}(0)=1$ for $\tau=0$ and $n_{\tau}(0)=0$ otherwise, equivalently, in terms of the Kronecker $\delta$-function, $n_{\tau}(0)=\delta_{0,\tau}$. First note that the binomial coefficient satisfies $\binom{x+1}{y+1}-\binom{x}{y}=\binom{x}{y+1}$ and $\binom{x}{y}=\frac{x-y+1}{y}\binom{x}{y-1}$, so we can write, for $t>\tau$
\begin{align}\label{eq1}
    \binom{t-\tau+\beta+1}{t-\tau+1}n_{\tau}(\tau)-\binom{t-\tau+\beta}{t-\tau}n_{\tau}(\tau)&=\binom{t-\tau+\beta}{t-\tau+1}n_{\tau}(\tau) \\
    &=\frac{t-\tau+\beta-t+\tau-1+1}{t-\tau+1}\binom{t-\tau+\beta}{t-\tau}n_{\tau}(\tau) \\ \label{eq1b}
     &=\frac{\beta}{t-\tau+1}\binom{t-\tau+\beta}{t-\tau}n_{\tau}(\tau).
\end{align}
Say $n_{\tau}(t)=\binom{t-\tau+\beta}{t-\tau}n_{\tau}(\tau)$, on the left-hand side of Equation \ref{eq1} we then recognize $\Delta_{t}n_{\tau}(t)=n_{\tau}(t+1)-n_{\tau}(t)$ and on the right-hand side of Equation \ref{eq1b} we recognize $ \beta n_{\tau}(t)/(t-\tau+1)$. Together, these correspond to the third line of Equation \ref{eqn:dynamics_a}. Given that $\binom{x}{y}=0$ for $y<0$, we also satisfy $n_{\tau}(0)=\delta_{0,\tau}$, thus automatically satisfying the first line of Equation \ref{eqn:dynamics_a}. Note this implies that $n_{\tau}(t)=0$ for all $t\leq \tau$. Finally, in order to satisfy the second line of Equation \ref{eqn:dynamics_a}, we require that 
\begin{align}
\alpha n(\tau-1)&=n_{\tau}(\tau)-n_{\tau}(\tau-1),\\
&=n_{\tau}(\tau). 
\end{align}
This in turn implies that 
\begin{align}
    n_{\tau}(\tau)&=\alpha \sum_{\tau'=0}^{\tau-1}n_{\tau'}(\tau-1), \\ \label{eqn:recurs}
    &=\alpha \sum_{\tau'=0}^{\tau-1}\binom{\tau-1-\tau'+\beta}{\tau-1-\tau'}n_{\tau'}(\tau'), 
\end{align}
which is a recursive relation involving $ n_{\tau}(\tau)$ only. To solve this relation we will use an identity and an ansatz. The identity we use comes from Concrete Mathematics (page 202) Equation 5.62 \citep{knuth1989concrete} and says that for each integer $n$, any real $r$,$s$,$t$, and summing over all $k$
\begin{equation}
    \sum_{k}\binom{tk+r}{k}\binom{tn-tk+s}{n-k}\frac{r}{tk+r}=\binom{tn+s+r}{n},
\end{equation}
which for $t=1$ reduces to
\begin{equation}
    \sum_{k}\binom{k+r-1}{k}\binom{n-k+s}{n-k}=\binom{n+s+r}{n}.
\end{equation}
Alternatively, if $\beta=r-1$, $n=\tau-1-u$ and $s=u\beta-1+u$, we can write it as
\begin{equation}
    \sum_{k}\binom{k+\beta}{k}\binom{\tau+u\beta-2-k}{\tau-1-u-k}=\binom{(u+1)\beta+\tau-1}{\tau-u-1}.
\end{equation}
Plugging in Equation \ref{eqn:recurs} the ansatz $ n_{\tau}(\tau)=\sum_{u=0}^{\tau}\binom{u\beta +\tau-1}{\tau-u}\alpha^{u}$ and using the earlier mentioned identity in the second to last step, we obtain for $\tau>0$ 
\begin{align}
    \sum_{u=0}^{\tau}\binom{u\beta +\tau-1}{\tau-u}\alpha^{u}
    &=\alpha \sum_{\tau'=0}^{\tau-1}\binom{\tau-1-\tau'+\beta}{\tau-1-\tau'}\sum_{u=0}^{\tau'}\binom{u\beta +\tau'-1}{\tau'-u}\alpha^{u}, \\
     &=\alpha \sum_{\tau'=0}^{\tau-1}\sum_{u=0}^{\tau-1}\binom{\tau-1-\tau'+\beta}{\tau-1-\tau'}\binom{u\beta +\tau'-1}{\tau'-u}\alpha^{u}, \\
       &=\alpha \sum_{u=0}^{\tau-1}\sum_{\tau'=0}^{\tau-1}\binom{\tau-1-\tau'+\beta}{\tau-1-\tau'}\binom{u\beta +\tau'-1}{\tau'-u}\alpha^{u}, \\
        &=\alpha \sum_{u=0}^{\tau-1}\left(\sum_{k=0}^{\tau-1}\binom{k+\beta}{k}\binom{u\beta +\tau-2-k}{\tau-1-u-k}\right)\alpha^{u}, \\
         &=\alpha \sum_{u=0}^{\tau-1}\binom{(u+1)\beta+\tau-1}{\tau-1-u}\alpha^{u},\\ \label{eqn:long}
         &= \sum_{u=1}^{\tau}\binom{u\beta+\tau-1}{\tau-u}\alpha^{u},
         \end{align}
The left- and right-hand side of Equation \ref{eqn:long} only disagree in case $\tau=0$, but since the ansatz satisfies $n_{0}(0)=\binom{-1}{0}=1$, by plugging in the ansatz in $n_{\tau}(t)=\binom{t-\tau+\beta}{t-\tau}n_{\tau}(\tau)$ we obtain the exact solution for all $\tau$ 
\begin{equation}\label{eqn:sol1_a}
    n_{\tau}(t)=\binom{t-\tau+\beta}{\beta} \sum_{u=0}^{\tau}\binom{u\beta +\tau-1}{\tau-u}\alpha^{u}.
\end{equation}
Since $n_{t}(t)=\alpha n(t-1)$, we have also derived an expression for $n(t)$ for $t\geq 1$, 
\begin{align}
    n(t-1)&=\frac{1}{\alpha}\sum_{u=0}^{t}\binom{u\beta +t-1}{t-u}\alpha^{u} \\
    \end{align}
    or, for $t\geq 0$
\begin{align}
    n(t)&= \frac{1}{\alpha}\sum_{u=0}^{t+1}\binom{u\beta +t}{t-u+1}\alpha^{u} \\
    &= \sum_{u=0}^{t+1}\binom{(u-1+1)\beta +t}{t-(u-1)}\alpha^{u-1}\\
    &= \sum_{u=-1}^{t}\binom{(u+1)\beta +t}{t-u}\alpha^{u} \\ \label{total_sum}
    &= \sum_{u=0}^{t}\binom{(u+1)\beta +t}{t-u}\alpha^{u} 
    \end{align}

\subsection*{Determine growth factor}

In Figure \ref{example_plot} we plot $n_{\tau}(t)$ for several $t$ on the left and $n(t)$ for several values of $\alpha$ and $\beta$ on the right. It is clear that for large $t$, the number of families increases exponentially, with a base tending to a constant $g$ that depends both on $\alpha$ and $\beta$. Let us first look at two extremes: say $\beta=0$, then summing Equation \ref{eqn:dynamics_a} left and right over all $\tau$ gives $\Delta_{t}n(t)=\alpha n(t)$ which directly suggest $n(t)\propto (1+\alpha)^{t}$. Let us consider another extreme where in the third line of Equation \ref{eqn:dynamics_a} we instead have the expression $\beta n_{\tau}(t)$, which is always larger than the original expression $\beta n_{\tau}(t)/(t-\tau+1)$ and therefore gives a larger estimation of $n(t)$. Again summing left and right over all $\tau$ would then give $\Delta_{t}n(t)=(\alpha+\beta) n(t)$ which directly suggests $n(t)\propto (1+\alpha+\beta)^{t}$. Therefore, we have that $1+\alpha\leq g <1+\alpha +\beta$.

To obtain a more precise expression for $g$, it is instructive to calculate the generating function $G(z)$ of $n(t)$, that is, $G(z)=\sum_{t=0}^{\infty}n(t)z^{t}$ for some $0\leq z<1$. Using the expression for $n(t)$ in Equation \ref{total_sum}, we obtain
\begin{align}
G(z)&=\sum_{t=0}^{\infty}\sum_{u=0}^{t}\binom{(u+1)\beta +t}{t-u}\alpha^{u}z^{t}, \\
&=\sum_{u=0}^{\infty}\sum_{t=0}^{\infty}\binom{(u+1)\beta +t+u}{t}\alpha^{u}z^{t+u}, \\
    &=\sum_{u=0}^{\infty}(\alpha z)^{u}\sum_{t=0}^{\infty}\binom{(u+1)\beta+t+u}{t}z^{t}, \\
     &=\sum_{u=0}^{\infty}(\alpha z)^{u}(1-z)^{-(1+\beta)(u+1)}\\
      &=\frac{1}{(1-z)^{1+\beta}}\sum_{u=0}^{\infty}\left(\frac{\alpha z}{(1-z)^{1+\beta}}\right)^{u}\\
      &=\frac{1}{(1-z)^{1+\beta}-\alpha z}
\end{align} 
We observe that $G(z)$ is only infinite when $(1-z)^{1+\beta}=\alpha z$. For real $0\leq z\leq 1$, $0\leq \alpha z\leq \alpha$ increases monotonically with $z$ and $1\geq (1-z)^{1+\beta}\geq 0$ decreases monotonically with $z$; and since $0<1$ and $\alpha>0$, there is one unique intersection point between these for $z=r$. Let me repeat the definition $G(z)=\sum_{t=0}^{\infty}n(t)z^{t}$ and suppose we can approximate $n(t)\simeq n_{0}g^t$ for some well chosen value $g$. For the right $g$, it counts $G(z)'=\sum_{t=0}^{\infty}n_{0}g^t z^{t}\approx G(z)$. For the approximation to count for any $z$, we require that $g=1/r$: if $g$ would be smaller than $1/r$, then $gr<1$ and $G(r)'$ would be a finite number where $G(r)$ is infinite; if $g$ would be larger than $1/r$, $G(z)'$ would be infinite for $z>r$ where $G(z)$ would be finite. $G(z)$ is therefore only well approximated by $G(z)'$ if $g=1/r$, in other words, a real number satisfying  
\begin{equation}\label{eqn:basis_a}
    (1-1/g)^{1+\beta}=\alpha/g. 
\end{equation}

Noting that $\alpha=g\left(1-\frac{1}{g}\right)^{1+\beta}$, we observe that increasing $\alpha$, keeping $\beta$ constant, necessarily implies increasing $g$ when $\beta$ is kept constant, and vice versa. We can also rewrite Equation \ref{eqn:basis_a} as
\begin{equation}\label{alternative}
    \beta=\frac{\log{g}-\log{(\alpha)}}{\log{g}-\log{(g-1)}}-1.
\end{equation}
As $g>\alpha$ and $g>1$, the numerator and denominator in this expression are always real and positive. When $\beta$ increases, the fraction needs to increase. Differentiating the denominator with respect to $g$ gives $1/g-1/(g-1)$, which is always negative for $g>1$. This implies that the denominator always decreases with increasing $g>1$. Keeping $\alpha$ constant, the numerator only increases with increasing $g$. To increase the fraction in Equation \ref{alternative} while keeping $\alpha$ constant, we necessarily need to increase $g$. 

\subsection*{The number of classifications per patent}
Finally, we calculate the number of classifications per family $W$ using that, upon introduction, each family has on average $W_{0}$ classifications. Assuming $W_{0}$ is constant, the number of unique patents is given by $\sum_{\tau}n_{\tau}(\tau)/W_{0}$, allowing us to write
\begin{align}
    W&=\frac{n(t)}{\sum_{\tau=0}^{t}n_{\tau}(\tau)/W_{0}}\\
    &=W_{0}\frac{n(t)}{\sum_{\tau=0}^{t}\alpha n(\tau-1)}\\
    &\approx W_{0}\frac{n_{0}g^{t}}{\sum_{\tau=0}^{t}\alpha n_{0}g^{\tau-1}}\\
      &\approx W_{0}\frac{g^{t}(g-1)}{\alpha (g^{t}-1)}\\
         &\approx W_{0}\frac{(g-1)}{\alpha}
\end{align}

\subsection*{Estimating values for parameters $\beta$ and $\alpha$}

In this section, I will explain in more detail how the values for parameters  $\beta$ and $\alpha$ are estimated. As explained in the main text, I am applying the model to 'all technology', meaning that $n_{\tau}(t)$ represents all classifications of patents with filing year $\tau$ at time $t$. For completeness, I repeat the data specifications: I consider patent families filed earliest between 1970 and 2019, with at least one US application and one in another jurisdiction. Using data from 4 Patstat editions (2013, 2016, 2019, and 2023), there are three distinct reclassification moments: 2013/2016,2016/2019 and 2019/2023. Reclassification happens yearly, so I aggregate the changes over approximately three years for each reclassification moment.  

\begin{figure}[tbhp]
\centering
\includegraphics[width=0.7\textwidth]{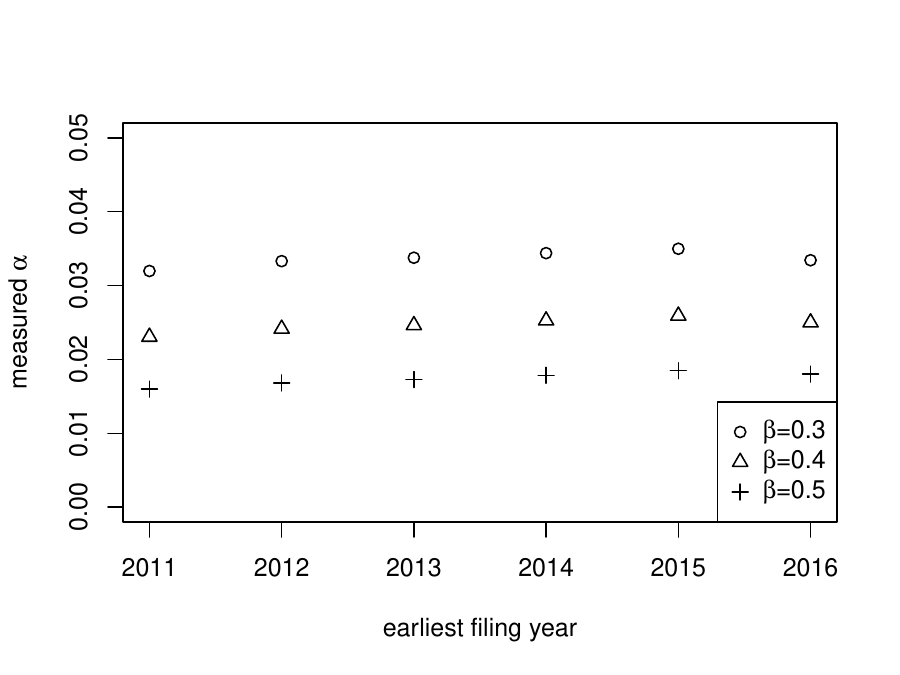}
\caption{Estimated values for $\alpha$ between 2011 and 2016 based the number of classifications of patents in the 2023 data and then subtracting the expected number of reclassifications over the years for various estimated values of $\beta$.}
\label{alpha_years}
\end{figure}

Estimation of $\beta$ (see Equation \ref{eqn:dynamics_a}) is straightforward given the fits in Fig. \ref{reclass_year} (in the main text): given these three reclassification moments, I estimate an average value $\beta\approx 0.4$. These plots are based on reclassifications on the subclass level only. However, the value $\beta$ are similar when considering different levels of classification. This is a result of using the net-reclassification per classification to determine beta, if the number of reclassifications per unique family were used, $\beta$ would depend on the level of classification used.

Next, Equation \ref{eqn:dynamics_a} indicates that $\alpha$ could be measured by $\Delta_{t}n_{\tau}(t)|_{\tau=t+1}/n(t)=n_{t+1}(t+1)/n(t)$, i.e. by dividing the number of classifications of newly introduced patents over the total number of classifications. Measuring this is challenging, however, because the number of newly introduced patents and classifications in recent years tends to be highly uncertain in patent databases, due to an (up to) 18-month period between the first filing and publication of a patent. For families with a much earlier filing year $Y$, however, the total number of classifications can reliably be measured; systematically reducing this number for each year between present and $Y$ (using an estimate for $\beta$) thus allows me to estimate how many classifications there were at year $Y$ (i.e., upon introduction). In other words, let me denote the number of classifications with filing year $Y$ at year $Y'$ by $C_{Y,Y'}$, then, given the present year $P$, the number of classifications upon introduction can be estimated by 
\begin{equation}\label{est_alpha}
    C_{Y,Y}=C_{Y,P}\left(1-\frac{\beta}{1} \right)\left(1-\frac{\beta}{2} \right)...\left(1-\frac{\beta}{P-Y} \right)
\end{equation}
Following \ref{eqn:dynamics_a}, $\alpha$ can be calculated dividing $C_{Y,Y}$ by the total number of classifications of patents with filing years $<Y$ at year $Y$, i.e. $\sum_{Y'}^{Y}C_{Y',Y}$. The $C_{Y',Y}$ can be estimated again using Equation \ref{est_alpha}, more specifically,   
\begin{equation}
    C_{Y',Y}=C_{Y,P}\left(1-\frac{\beta}{Y-Y'+1} \right)\left(1-\frac{\beta}{Y-Y'+2} \right)...\left(1-\frac{\beta}{Y'+P} \right)
\end{equation}
As this relation shows, the effect of transforming very early years (i.e. $Y'<<Y$) is limited as the fraction $\frac{\beta}{Y-Y'+1}$ is then very small. Therefore, when estimating $C_{Y,Y}$ in the following, I only transform the years $Y'$ for $Y'\geq Y-10$. 
The results of determining $C_{Y,Y}$ for $2011, 2012, 2013, 2014, 2015,$ and $2016$ for three estimated values of $\beta$ are included in Fig. \ref{alpha_years}. Taking 2023 as present day, the values for $C_{Y,2023}$ are based on the 2023 dataset. The results indicate that the estimated values for $\alpha$ are rather stable of the years, justifying approximating it to be constant. Unsurprisingly, for greater $\beta$, the effect of reducing classifications is greater, therefore leading to smaller $C_{Y,Y}$, therefore leading to smaller estimates of $\alpha$. Using the estimated value for $\beta\approx 0.4$ and classifications on the CPC group level, I estimate $\alpha$ to be between $0.024$ and $0.027$ (Fig. \ref{alpha_years}), on the CPC subclass level, these values are slightly lower, between $0.021$ and $0.023$. Finally, the estimates for $C_{Y,Y}$ allow me to directly estimate $W_{0}$ i.e. is the number of the number of patents classifications per family upon introduction, by dividing the $C_{Y,Y}$ by the unique number of families in $Y$. For the years plotted in Fig. \ref{alpha_years}, I find that $1<W_{0}<1.5$ on the CPC group level and $0.6<W_{0}<0.8$ on the CPC subclass level. While values for $W_{0}<1$ may seem counter-intuitive, it should be stressed these estimates are based on a chosen, fixed value of $\beta$, which may not be a realistic assumption for the years considered.  

\subsection*{Log-log plots empirical analysis}

\begin{figure}[tbhp]
\centering
\includegraphics[width=\textwidth]{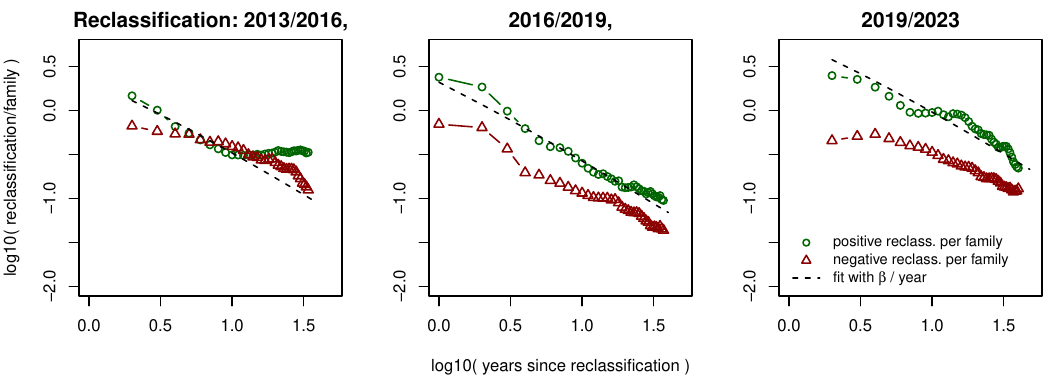}
\caption{The number of positive and negative reclassifications per family per filing year on a log log plot, note the horizontal axis ('years since reclassification') is reversed in comparison to Fig. \ref{reclass_year}. The slopes of the fits (included here for the positive reclassifications only) is $-1$, corresponding to an inverse relation with the time since reclassification.}
\label{log_years_pos_neg}
\end{figure}

In line with Figure \ref{reclass_year}, we again plot the number of reclassifications per year in Figure \ref{log_years_pos_neg}, except that we separately plot the negative and positive reclassifications, the horizontal axis is reversed, and both axes are log-transformed. Because the horizontal axes are reversed, the sharp increase towards the asymptote in Figure \ref{reclass_year} is instead a sharp decline in Fig. \ref{log_years_pos_neg}, observable for both the negative and positive reclassifications. The number of classifications added tend to be larger than those deleted, except for the reclassification moment between 2013 and 2016 (the reasons for this remain unclear). 

Furthermore, in Fig. \ref{log_years_pos_neg} we include fits of the number of classifications added based on a simple inverse proportional relation with the number of years since reclassification. This relation, as any power law, appears as a straight line in this log-log plot. As the slope of the line is determined by the exponent (here $-1$). The data appear to follow the same slope, which suggests that the number of reclassifications are approximately inversely proportional with the number of years since reclassification.   

\subsection*{Robustness checks and OLS results}

\begin{figure}[tbhp]
\centering
\includegraphics[width=\textwidth]{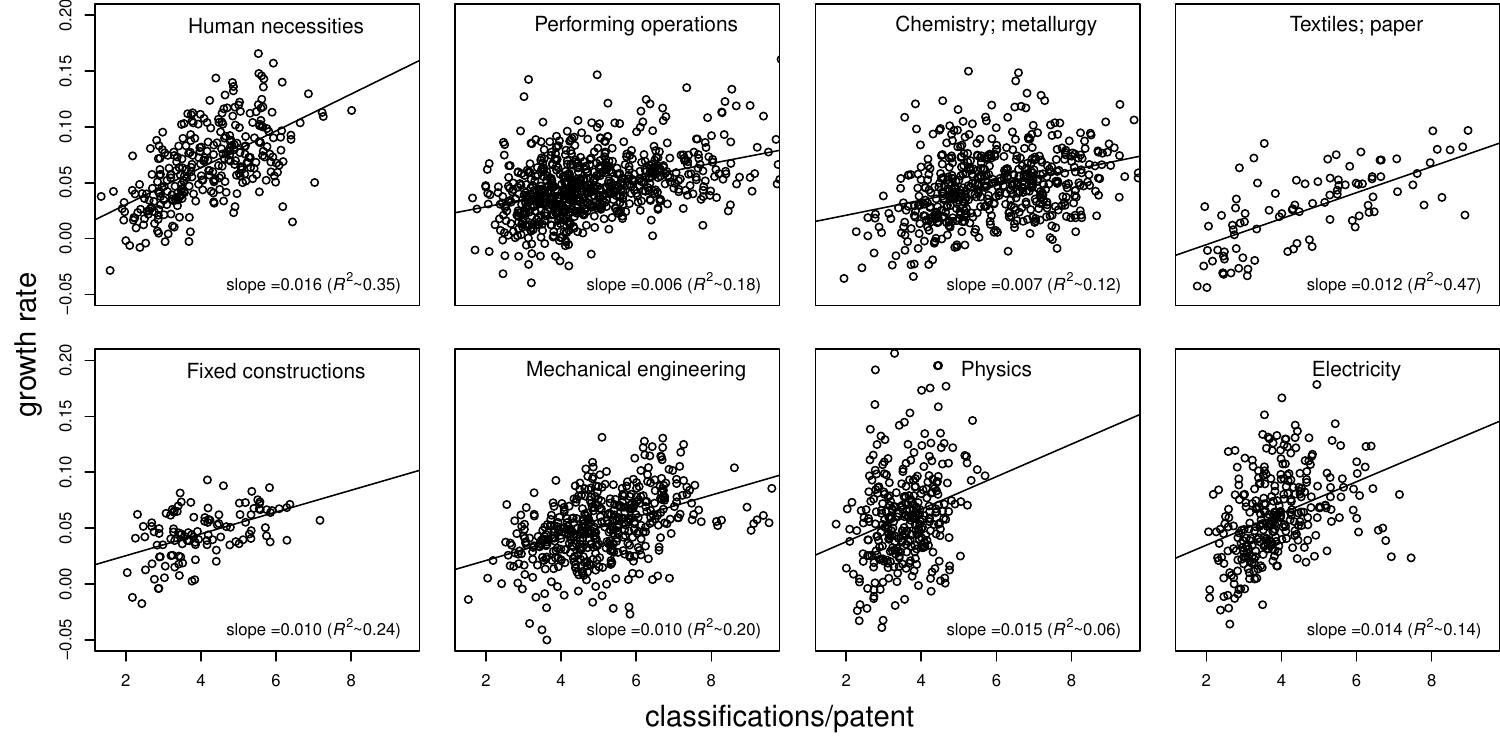}
\caption{Growth rate $g_{k}-1$ and number of classifications per patent $w_{k}$ for each CPC group $k$, except the groups in subclasses C10M, C10N, B32B, and F17C.}
\label{growth_class_all}
\end{figure}

Let me start with Fig. \ref{growth_class_all}, where I plot the growth rates $g_{k}-1$ for the number of classifications per patent $w_{k}$ (on the CPC group level) for each group $k$ similar to the plot in Fig. \ref{sections_growth} in the main text, except I leave out the groups in subclasses C10M, C10N, B32B, and F17C, which have exceptionally large $w_{k}>9$. 
Compared to the results in the main text, the correlations as measured by $R^2$ in for the sections Performing Operations, Chemistry/metallurgy and Mechanical engineering are much higher, also visually the fits are much better. As the omitted subclasses have relatively large $w_{k}$, this could suggest that especially for lower $w_{k}$, the relation between $g_{k}-1$ and $w_{k}$ applies and might taper off for higher values. However, little can be said about this relation as the number of groups in C10M, C10N, B32B, and F17C are only a very small fraction of the total. In the following robustness checks, I focus on the relations without these subclasses, however, the results are still highly significant when including them.  

Finally, I presents the details of three robustness checks I did on the observed positive relation between average classifications per patent $W_{k}$ and the yearly growth rate $g_{k}-1$ of a CPC group $k$. As the strength of the relation may be rather different for different CPC sections, I analyze them separately \footnote{I do not include the results for CPC section Y, which is a CPC wide tagging scheme to signal the potential of a technology to reduce greenhouse gas emissions and adapt to climate change. The Y classifications are however counted in the number of classifications per patent.}. We anticipate three criticisms: 
\begin{enumerate}
    \item That the observed relation between $W_{k}$ and  $g_{k}-1$ may only be an indirect effect because both quantities would correlate the size of CPC group $k$ (in number of patents).
    \item That the observed relation between $W_{k}$ and  $g_{k}-1$ may only be an indirect effect because both quantities correlate to the characteristic to have many recent patents (this would count for $W_{k}$ because reclassification especially targets recent patents, and many recent patents would bias $g_{k}-1$ to greater values). 
    \item That the observed relation between $W_{k}$ and  $g_{k}-1$ is only an 'apparent effect' because more reclassifications results in more classifications but not necessarily in more patents (or faster growth). This criticism suggests that if the $g_{k}-1$ would be based not on the number of classifications but on the number of unique patents, there would not be a relation with $W_{k}$.  
    \end{enumerate}
In the first robustness check, I address the first two criticisms by accounting for the (log) total group size $n_{k}$ and the (log) number of patents in three recent years for each group in estimating a linear model using an Ordinary Least Squares (OLS). For the recent years I choose 2014, 2013 and 2012, as for any later year I am dealing with the apparent decline as discussed in the main text. The estimated relation is therefore summarized as
\begin{align}\nonumber
     growth\_rate \propto & constant+ class\_per\_family + log\_group\_total \\ 
     & + log\_patents\_2014 + log\_patents\_2013 + log\_patents\_2012
   \end{align}
The results are presented in Tables \ref{A_av} and \ref{B_av}. From these tables it is clear that the relation between $W_{k}$ and  $g_{k}-1$ is still robust after including the total number of patents in each group and the number of patents in recent years. Furthermore, we conclude that for most CPC sections, either the (log) number of patents in those recent years are not significant, or result in the fact the log group total number of patents are not (or no longer) significant. 

\begin{table}[!htbp] 
\small
\centering 
  \caption{OLS results classification per family and number of patents in recent years (A)} 
  \parbox{.45\linewidth}{
  \subcaption*{Section A}
  \label{A_av} 
\begin{tabular}{@{\extracolsep{5pt}}lc} 
\\[-1.8ex]\hline 
\hline \\[-1.8ex] 
 & \multicolumn{1}{c}{\textit{Dependent variable:}} \\ 
\cline{2-2} 
\\[-1.8ex] & growth\_rate \\ 
\hline \\[-1.8ex] 
 class\_per\_family & 0.010$^{***}$ \\ 
  & (0.001) \\ 
  & \\ 
 log\_group\_total & $-$0.014$^{***}$ \\ 
  & (0.003) \\ 
  & \\ 
 log\_patents\_2014 & 0.006 \\ 
  & (0.004) \\ 
  & \\ 
 log\_patents\_2013 & 0.015$^{***}$ \\ 
  & (0.005) \\ 
  & \\ 
 log\_patents\_2012 & 0.008$^{*}$ \\ 
  & (0.004) \\ 
  & \\ 
 Constant & 0.006 \\ 
  & (0.014) \\ 
  & \\ 
\hline \\[-1.8ex] 
Observations & 300 \\ 
R$^{2}$ & 0.728 \\ 
Adjusted R$^{2}$ & 0.723 \\ 
Residual Std. Error & 0.018 (df = 294) \\ 
F Statistic & 157.421$^{***}$ (df = 5; 294) \\ 
\hline 
\hline \\[-1.8ex] 
\end{tabular} 
\bigskip
 \subcaption*{Section C}
\begin{tabular}{@{\extracolsep{5pt}}lc} 
\\[-1.8ex]\hline 
\hline \\[-1.8ex] 
 & \multicolumn{1}{c}{\textit{Dependent variable:}} \\ 
\cline{2-2} 
\\[-1.8ex] & growth\_rate \\ 
\hline \\[-1.8ex] 
 class\_per\_family & 0.003$^{***}$ \\ 
  & (0.001) \\ 
  & \\ 
 log\_group\_total & $-$0.020$^{***}$ \\ 
  & (0.002) \\ 
  & \\ 
 log\_patents\_2014 & 0.010$^{***}$ \\ 
  & (0.003) \\ 
  & \\ 
 log\_patents\_2013 & 0.010$^{***}$ \\ 
  & (0.004) \\ 
  & \\ 
 log\_patents\_2012 & 0.011$^{***}$ \\ 
  & (0.003) \\ 
  & \\ 
 Constant & 0.043$^{***}$ \\ 
  & (0.011) \\ 
  & \\ 
\hline \\[-1.8ex] 
Observations & 543 \\ 
R$^{2}$ & 0.556 \\ 
Adjusted R$^{2}$ & 0.552 \\ 
Residual Std. Error & 0.020 (df = 537) \\ 
F Statistic & 134.303$^{***}$ (df = 5; 537) \\ 
\hline 
\hline \\[-1.8ex] 
\end{tabular} 
}
\hfill
  \parbox{.45\linewidth}{
  \subcaption*{Section B}
\begin{tabular}{@{\extracolsep{5pt}}lc} 
\\[-1.8ex]\hline 
\hline \\[-1.8ex] 
 & \multicolumn{1}{c}{\textit{Dependent variable:}} \\ 
\cline{2-2} 
\\[-1.8ex] & growth\_rate \\ 
\hline \\[-1.8ex] 
 class\_per\_family & 0.003$^{***}$ \\ 
  & (0.0004) \\ 
  & \\ 
 log\_group\_total & $-$0.019$^{***}$ \\ 
  & (0.002) \\ 
  & \\ 
 log\_patents\_2014 & 0.005$^{**}$ \\ 
  & (0.002) \\ 
  & \\ 
 log\_patents\_2013 & 0.009$^{***}$ \\ 
  & (0.003) \\ 
  & \\ 
 log\_patents\_2012 & 0.014$^{***}$ \\ 
  & (0.002) \\ 
  & \\ 
 Constant & 0.052$^{***}$ \\ 
  & (0.009) \\ 
  & \\ 
\hline \\[-1.8ex] 
Observations & 752 \\ 
R$^{2}$ & 0.532 \\ 
Adjusted R$^{2}$ & 0.529 \\ 
Residual Std. Error & 0.019 (df = 746) \\ 
F Statistic & 169.446$^{***}$ (df = 5; 746) \\ 
\hline 
\hline \\[-1.8ex] 
\end{tabular} 
\bigskip
 \subcaption*{Section D}
\begin{tabular}{@{\extracolsep{5pt}}lc} 
\\[-1.8ex]\hline 
\hline \\[-1.8ex] 
 & \multicolumn{1}{c}{\textit{Dependent variable:}} \\ 
\cline{2-2} 
\\[-1.8ex] & growth\_rate \\ 
\hline \\[-1.8ex] 
 class\_per\_family & 0.003$^{***}$ \\ 
  & (0.001) \\ 
  & \\ 
 log\_group\_total & $-$0.027$^{***}$ \\ 
  & (0.005) \\ 
  & \\ 
 log\_patents\_2014 & 0.021$^{***}$ \\ 
  & (0.005) \\ 
  & \\ 
 log\_patents\_2013 & 0.001 \\ 
  & (0.005) \\ 
  & \\ 
 log\_patents\_2012 & 0.013$^{***}$ \\ 
  & (0.004) \\ 
  & \\ 
 Constant & 0.078$^{***}$ \\ 
  & (0.026) \\ 
  & \\ 
\hline \\[-1.8ex] 
Observations & 94 \\ 
R$^{2}$ & 0.793 \\ 
Adjusted R$^{2}$ & 0.782 \\ 
Residual Std. Error & 0.016 (df = 88) \\ 
F Statistic & 67.615$^{***}$ (df = 5; 88) \\ 
\hline 
\hline \\[-1.8ex] 
\end{tabular} 
}
\begin{tabular}{@{\extracolsep{5pt}}lc} 
\hline
\textit{Note:}  & \multicolumn{1}{r}{$^{*}$p$<$0.1; $^{**}$p$<$0.05; $^{***}$p$<$0.01} \\ 
\end{tabular} 
\end{table} 

\begin{table}[!htbp] 
\small
\centering 
  \caption{OLS results classification per family and patent numbers in recent years (B)} 
  \label{B_av}
  \parbox{.45\linewidth}{
 \subcaption*{Section E}
\begin{tabular}{@{\extracolsep{5pt}}lc} 
\\[-1.8ex]\hline 
\hline \\[-1.8ex] 
 & \multicolumn{1}{c}{\textit{Dependent variable:}} \\ 
\cline{2-2} 
\\[-1.8ex] & growth\_rate \\ 
\hline \\[-1.8ex] 
 class\_per\_family & 0.008$^{***}$ \\ 
  & (0.001) \\ 
  & \\ 
 log\_group\_total & $-$0.023$^{***}$ \\ 
  & (0.006) \\ 
  & \\ 
 log\_patents\_2014 & 0.021$^{***}$ \\ 
  & (0.007) \\ 
  & \\ 
 log\_patents\_2013 & 0.009 \\ 
  & (0.006) \\ 
  & \\ 
 log\_patents\_2012 & $-$0.002 \\ 
  & (0.006) \\ 
  & \\ 
 Constant & 0.063$^{**}$ \\ 
  & (0.024) \\ 
  & \\ 
\hline \\[-1.8ex] 
Observations & 114 \\ 
R$^{2}$ & 0.560 \\ 
Adjusted R$^{2}$ & 0.540 \\ 
Residual Std. Error & 0.015 (df = 108) \\ 
F Statistic & 27.511$^{***}$ (df = 5; 108) \\ 
\hline 
\hline \\[-1.8ex] 
\end{tabular} 
\bigskip
  \subcaption*{Section G}
\begin{tabular}{@{\extracolsep{5pt}}lc} 
\\[-1.8ex]\hline 
\hline \\[-1.8ex] 
 & \multicolumn{1}{c}{\textit{Dependent variable:}} \\ 
\cline{2-2} 
\\[-1.8ex] & growth\_rate \\ 
\hline \\[-1.8ex] 
 class\_per\_family & 0.009$^{***}$ \\ 
  & (0.002) \\ 
  & \\ 
 log\_group\_total & $-$0.001 \\ 
  & (0.003) \\ 
  & \\ 
 log\_patents\_2014 & 0.003 \\ 
  & (0.006) \\ 
  & \\ 
 log\_patents\_2013 & 0.014$^{**}$ \\ 
  & (0.007) \\ 
  & \\ 
 log\_patents\_2012 & 0.005 \\ 
  & (0.006) \\ 
  & \\ 
 Constant & $-$0.065$^{***}$ \\ 
  & (0.016) \\ 
  & \\ 
\hline \\[-1.8ex] 
Observations & 317 \\ 
R$^{2}$ & 0.584 \\ 
Adjusted R$^{2}$ & 0.577 \\ 
Residual Std. Error & 0.027 (df = 311) \\ 
F Statistic & 87.214$^{***}$ (df = 5; 311) \\ 
\hline 
\hline \\[-1.8ex] 
\end{tabular} 
}
\hfill
  \parbox{.45\linewidth}{
    \subcaption*{Section F}
  \begin{tabular}{@{\extracolsep{5pt}}lc} 
\\[-1.8ex]\hline 
\hline \\[-1.8ex] 
 & \multicolumn{1}{c}{\textit{Dependent variable:}} \\ 
\cline{2-2} 
\\[-1.8ex] & growth\_rate \\ 
\hline \\[-1.8ex] 
 class\_per\_family & 0.006$^{***}$ \\ 
  & (0.001) \\ 
  & \\ 
 log\_group\_total & $-$0.023$^{***}$ \\ 
  & (0.002) \\ 
  & \\ 
 log\_patents\_2014 & 0.012$^{***}$ \\ 
  & (0.003) \\ 
  & \\ 
 log\_patents\_2013 & 0.008$^{**}$ \\ 
  & (0.004) \\ 
  & \\ 
 log\_patents\_2012 & 0.011$^{***}$ \\ 
  & (0.004) \\ 
  & \\ 
 Constant & 0.047$^{***}$ \\ 
  & (0.011) \\ 
  & \\ 
\hline \\[-1.8ex] 
Observations & 478 \\ 
R$^{2}$ & 0.618 \\ 
Adjusted R$^{2}$ & 0.614 \\ 
Residual Std. Error & 0.018 (df = 472) \\ 
F Statistic & 152.855$^{***}$ (df = 5; 472) \\ 
\hline 
\hline \\[-1.8ex] 
\end{tabular} 
\bigskip
  \subcaption*{Section H}
\begin{tabular}{@{\extracolsep{5pt}}lc} 
\\[-1.8ex]\hline 
\hline \\[-1.8ex] 
 & \multicolumn{1}{c}{\textit{Dependent variable:}} \\ 
\cline{2-2} 
\\[-1.8ex] & growth\_rate \\ 
\hline \\[-1.8ex] 
 class\_per\_family & 0.006$^{***}$ \\ 
  & (0.001) \\ 
  & \\ 
 log\_group\_total & $-$0.010$^{***}$ \\ 
  & (0.003) \\ 
  & \\ 
 log\_patents\_2014 & 0.009$^{*}$ \\ 
  & (0.005) \\ 
  & \\ 
 log\_patents\_2013 & 0.004 \\ 
  & (0.005) \\ 
  & \\ 
 log\_patents\_2012 & 0.012$^{**}$ \\ 
  & (0.005) \\ 
  & \\ 
 Constant & $-$0.003 \\ 
  & (0.014) \\ 
  & \\ 
\hline \\[-1.8ex] 
Observations & 307 \\ 
R$^{2}$ & 0.645 \\ 
Adjusted R$^{2}$ & 0.639 \\ 
Residual Std. Error & 0.022 (df = 301) \\ 
F Statistic & 109.333$^{***}$ (df = 5; 301) \\ 
\hline 
\hline \\[-1.8ex] 
\end{tabular} 
}
\begin{tabular}{@{\extracolsep{5pt}}lc} 
\hline
\textit{Note:}  & \multicolumn{1}{r}{$^{*}$p$<$0.1; $^{**}$p$<$0.05; $^{***}$p$<$0.01} \\ 
\end{tabular} 
\end{table} 

In the second robustness check, I account for the second criticism in a different way. Instead of including the number of patents in the recent years as variables, we can also redefine $W_{k}$ to make it 'less biased' to the number of patents in recent years: as there are relatively many reclassification in recent patents, if a group has relatively many recent patents, its average number of classifications per patent might be positively biased. To sidestep this, I slightly redefine $W_{k}$ as the overall average of the yearly average of classification per patent (for each group $k$), which I will refer to as the 'year averaged classification per family' or 'year\_av\_class\_per\_family'. The second criticism would be justified if I would no longer find a significant relation between $W_{k}$ and  $g_{k}-1$ after redefining $W_{k}$. The tested relation then becomes   
\begin{equation}
     \textit{growth\_rate $\propto$ constant+ year\_av\_class\_per\_family + log\_group\_total },
   \end{equation}
where for completeness I again account for group size. In Tables \ref{A_yearav} and \ref{B_yearav} the outcomes of the estimations are presented. I observe that there remains a highly significant positive relation between the growth rates and year\_av\_class\_per\_family, so even when I account for the bias brought to $W_{k}$ by the number of recent patents, there is still a significant relation with the growth rates.   
\begin{table}[!htbp] 
\small
\centering 
  \caption{OLS results with year-averaged classification per family (A)} 
  \parbox{.45\linewidth}{
  \subcaption*{Section A}
  \label{A_yearav} 
\begin{tabular}{@{\extracolsep{5pt}}lc} 
\\[-1.8ex]\hline 
\hline \\[-1.8ex] 
 & \multicolumn{1}{c}{\textit{Dependent variable:}} \\ 
\cline{2-2} 
\\[-1.8ex] & growth\_rate \\ 
\hline \\[-1.8ex] 
 year\_av\_class\_per\_family & 0.011$^{***}$ \\ 
  & (0.001) \\ 
  & \\ 
 log\_group\_total & 0.017$^{***}$ \\ 
  & (0.001) \\ 
  & \\ 
 Constant & $-$0.100$^{***}$ \\ 
  & (0.008) \\ 
  & \\ 
\hline \\[-1.8ex] 
Observations & 300 \\ 
R$^{2}$ & 0.598 \\ 
Adjusted R$^{2}$ & 0.595 \\ 
Residual Std. Error & 0.022 (df = 297) \\ 
F Statistic & 220.925$^{***}$ (df = 2; 297) \\ 
\hline 
\hline \\[-1.8ex] 
\end{tabular} 
\bigskip
 \subcaption*{Section C}
\begin{tabular}{@{\extracolsep{5pt}}lc} 
\\[-1.8ex]\hline 
\hline \\[-1.8ex] 
 & \multicolumn{1}{c}{\textit{Dependent variable:}} \\ 
\cline{2-2} 
\\[-1.8ex] & growth\_rate \\ 
\hline \\[-1.8ex] 
 year\_av\_class\_per\_family & 0.004$^{***}$ \\ 
  & (0.001) \\ 
  & \\ 
 log\_group\_total & 0.015$^{***}$ \\ 
  & (0.001) \\ 
  & \\ 
 Constant & $-$0.085$^{***}$ \\ 
  & (0.009) \\ 
  & \\ 
\hline \\[-1.8ex] 
Observations & 543 \\ 
R$^{2}$ & 0.289 \\ 
Adjusted R$^{2}$ & 0.287 \\ 
Residual Std. Error & 0.026 (df = 540) \\ 
F Statistic & 109.954$^{***}$ (df = 2; 540) \\ 
\hline 
\hline \\[-1.8ex] 
\end{tabular} 
}
\hfill
  \parbox{.45\linewidth}{
  \subcaption*{Section B}
\begin{tabular}{@{\extracolsep{5pt}}lc} 
\\[-1.8ex]\hline 
\hline \\[-1.8ex] 
 & \multicolumn{1}{c}{\textit{Dependent variable:}} \\ 
\cline{2-2} 
\\[-1.8ex] & growth\_rate \\ 
\hline \\[-1.8ex] 
 year\_av\_class\_per\_family & 0.004$^{***}$ \\ 
  & (0.001) \\ 
  & \\ 
 log\_group\_total & 0.014$^{***}$ \\ 
  & (0.001) \\ 
  & \\ 
 Constant & $-$0.069$^{***}$ \\ 
  & (0.006) \\ 
  & \\ 
\hline \\[-1.8ex] 
Observations & 752 \\ 
R$^{2}$ & 0.336 \\ 
Adjusted R$^{2}$ & 0.334 \\ 
Residual Std. Error & 0.022 (df = 749) \\ 
F Statistic & 189.633$^{***}$ (df = 2; 749) \\ 
\hline 
\hline \\[-1.8ex] 

\end{tabular} 
\bigskip
 \subcaption*{Section D}
\begin{tabular}{@{\extracolsep{5pt}}lc} 
\\[-1.8ex]\hline 
\hline \\[-1.8ex] 
 & \multicolumn{1}{c}{\textit{Dependent variable:}} \\ 
\cline{2-2} 
\\[-1.8ex] & growth\_rate \\ 
\hline \\[-1.8ex] 
 year\_av\_class\_per\_family & 0.011$^{***}$ \\ 
  & (0.001) \\ 
  & \\ 
 log\_group\_total & 0.015$^{***}$ \\ 
  & (0.003) \\ 
  & \\ 
 Constant & $-$0.120$^{***}$ \\ 
  & (0.023) \\ 
  & \\ 
\hline \\[-1.8ex] 
Observations & 94 \\ 
R$^{2}$ & 0.490 \\ 
Adjusted R$^{2}$ & 0.479 \\ 
Residual Std. Error & 0.025 (df = 91) \\ 
F Statistic & 43.778$^{***}$ (df = 2; 91) \\ 
\hline 
\hline \\[-1.8ex] 
\end{tabular} 
}
\begin{tabular}{@{\extracolsep{5pt}}lc} 
\hline
\textit{Note:}  & \multicolumn{1}{r}{$^{*}$p$<$0.1; $^{**}$p$<$0.05; $^{***}$p$<$0.01} \\ 
\end{tabular} 
\end{table} 

\begin{table}[!htbp] 
\small
\centering 
  \caption{OLS results with year-averaged classification per family (B)} 
  \label{B_yearav}
  \parbox{.45\linewidth}{
 \subcaption*{Section E}
\begin{tabular}{@{\extracolsep{5pt}}lc} 
\\[-1.8ex]\hline 
\hline \\[-1.8ex] 
 & \multicolumn{1}{c}{\textit{Dependent variable:}} \\ 
\cline{2-2} 
\\[-1.8ex] & growth\_rate \\ 
\hline \\[-1.8ex] 
 year\_av\_class\_per\_family & 0.008$^{***}$ \\ 
  & (0.002) \\ 
  & \\ 
 log\_group\_total & 0.011$^{***}$ \\ 
  & (0.002) \\ 
  & \\ 
 Constant & $-$0.060$^{***}$ \\ 
  & (0.015) \\ 
  & \\ 
\hline \\[-1.8ex] 
Observations & 114 \\ 
R$^{2}$ & 0.360 \\ 
Adjusted R$^{2}$ & 0.349 \\ 
Residual Std. Error & 0.018 (df = 111) \\ 
F Statistic & 31.264$^{***}$ (df = 2; 111) \\ 
\hline 
\hline \\[-1.8ex] 
\end{tabular} 
\bigskip
  \subcaption*{Section G}
\begin{tabular}{@{\extracolsep{5pt}}lc} 
\\[-1.8ex]\hline 
\hline \\[-1.8ex] 
 & \multicolumn{1}{c}{\textit{Dependent variable:}} \\ 
\cline{2-2} 
\\[-1.8ex] & growth\_rate \\ 
\hline \\[-1.8ex] 
 year\_av\_class\_per\_family & 0.009$^{***}$ \\ 
  & (0.003) \\ 
  & \\ 
 log\_group\_total & 0.022$^{***}$ \\ 
  & (0.001) \\ 
  & \\ 
 Constant & $-$0.137$^{***}$ \\ 
  & (0.015) \\ 
  & \\ 
\hline \\[-1.8ex] 
Observations & 317 \\ 
R$^{2}$ & 0.424 \\ 
Adjusted R$^{2}$ & 0.420 \\ 
Residual Std. Error & 0.032 (df = 314) \\ 
F Statistic & 115.556$^{***}$ (df = 2; 314) \\ 
\hline 
\hline \\[-1.8ex] 
\end{tabular} 
}
\hfill
  \parbox{.45\linewidth}{
 \subcaption*{Section F}
\begin{tabular}{@{\extracolsep{5pt}}lc} 
\\[-1.8ex]\hline 
\hline \\[-1.8ex] 
 & \multicolumn{1}{c}{\textit{Dependent variable:}} \\ 
\cline{2-2} 
\\[-1.8ex] & growth\_rate \\ 
\hline \\[-1.8ex] 
 year\_av\_class\_per\_family & 0.006$^{***}$ \\ 
  & (0.001) \\ 
  & \\ 
 log\_group\_total & 0.017$^{***}$ \\ 
  & (0.001) \\ 
  & \\ 
 Constant & $-$0.093$^{***}$ \\ 
  & (0.011) \\ 
  & \\ 
\hline \\[-1.8ex] 
Observations & 478 \\ 
R$^{2}$ & 0.282 \\ 
Adjusted R$^{2}$ & 0.279 \\ 
Residual Std. Error & 0.025 (df = 475) \\ 
F Statistic & 93.329$^{***}$ (df = 2; 475) \\ 
\hline 
\hline \\[-1.8ex] 
\end{tabular} 
\bigskip
  \subcaption*{Section H}
\begin{tabular}{@{\extracolsep{5pt}}lc} 
\\[-1.8ex]\hline 
\hline \\[-1.8ex] 
 & \multicolumn{1}{c}{\textit{Dependent variable:}} \\ 
\cline{2-2} 
\\[-1.8ex] & growth\_rate \\ 
\hline \\[-1.8ex] 
 year\_av\_class\_per\_family & 0.008$^{***}$ \\ 
  & (0.002) \\ 
  & \\ 
 log\_group\_total & 0.018$^{***}$ \\ 
  & (0.001) \\ 
  & \\ 
 Constant & $-$0.110$^{***}$ \\ 
  & (0.011) \\ 
  & \\ 
\hline \\[-1.8ex] 
Observations & 307 \\ 
R$^{2}$ & 0.458 \\ 
Adjusted R$^{2}$ & 0.454 \\ 
Residual Std. Error & 0.027 (df = 304) \\ 
F Statistic & 128.196$^{***}$ (df = 2; 304) \\ 
\hline 
\hline \\[-1.8ex] 
\end{tabular} 
}
\begin{tabular}{@{\extracolsep{5pt}}lc} 
\hline
\textit{Note:}  & \multicolumn{1}{r}{$^{*}$p$<$0.1; $^{**}$p$<$0.05; $^{***}$p$<$0.01} \\ 
\end{tabular} 
\end{table} 

In the third and last robustness check, I address the earlier mentioned third criticism. According to that criticism, the boosting of growth rates of groups with many reclassified patents is not a real acceleration because the reclassified patents are only 'copies', i.e. they already exist elsewhere in another group. To counter this criticism, I slightly redefine the growth rate (and group totals) such that it only counts each unique patent once. If a patent is classified in multiple groups, I divide it evenly over these groups, i.e. I do a fractional attribution of patents across groups. I will refer to resulting growth rate and group totals as 'growth\_rate\_fractional' and 'group\_total\_fractional'. The third criticism would be justified if I would no longer find a significant relation between $W_{k}$ and $g_{k}-1$ after redefining the growth rate as such. The tested relation therefor becomes 
\begin{equation}
     \textit{growth\_rate $\propto$ constant+ year\_av\_class\_per\_family + log\_group\_total },
   \end{equation}
where for completeness I again account for group size. In Tables \ref{A_frac} and \ref{B_frac} the outcomes of the estimations are presented. I observe that there remains a highly significant positive relation between the redefined growth rates and classification per family.   

\begin{table}[!htbp] 
\small
\centering 
  \caption{OLS results with fractional patent counting (A)} 
  \parbox{.45\linewidth}{
  \subcaption*{Section A}
  \label{A_frac} 
\begin{tabular}{@{\extracolsep{5pt}}lc} 
\\[-1.8ex]\hline 
\hline \\[-1.8ex] 
 & \multicolumn{1}{c}{\textit{Dependent variable:}} \\ 
\cline{2-2} 
\\[-1.8ex] & growth\_rate\_fractional \\ 
\hline \\[-1.8ex] 
 class\_per\_family & 0.015$^{***}$ \\ 
  & (0.001) \\ 
  & \\ 
 log\_group\_total\_fractional & 0.018$^{***}$ \\ 
  & (0.001) \\ 
  & \\ 
 Constant & $-$0.118$^{***}$ \\ 
  & (0.009) \\ 
  & \\ 
\hline \\[-1.8ex] 
Observations & 300 \\ 
R$^{2}$ & 0.573 \\ 
Adjusted R$^{2}$ & 0.570 \\ 
Residual Std. Error & 0.023 (df = 297) \\ 
F Statistic & 198.875$^{***}$ (df = 2; 297) \\ 
\hline 
\hline \\[-1.8ex] 
\end{tabular} 
\bigskip
 \subcaption*{Section C}
\begin{tabular}{@{\extracolsep{5pt}}lc} 
\\[-1.8ex]\hline 
\hline \\[-1.8ex] 
 & \multicolumn{1}{c}{\textit{Dependent variable:}} \\ 
\cline{2-2} 
\\[-1.8ex] & growth\_rate\_fractional \\ 
\hline \\[-1.8ex] 
 class\_per\_family & 0.008$^{***}$ \\ 
  & (0.001) \\ 
  & \\ 
 log\_group\_total\_fractional & 0.013$^{***}$ \\ 
  & (0.001) \\ 
  & \\ 
 Constant & $-$0.091$^{***}$ \\ 
  & (0.010) \\ 
  & \\ 
\hline \\[-1.8ex] 
Observations & 543 \\ 
R$^{2}$ & 0.240 \\ 
Adjusted R$^{2}$ & 0.237 \\ 
Residual Std. Error & 0.028 (df = 540) \\ 
F Statistic & 85.203$^{***}$ (df = 2; 540) \\ 
\hline 
\hline \\[-1.8ex] 
\end{tabular} 
}
\hfill
  \parbox{.45\linewidth}{
  \subcaption*{Section B}
\begin{tabular}{@{\extracolsep{5pt}}lc} 
\\[-1.8ex]\hline 
\hline \\[-1.8ex] 
 & \multicolumn{1}{c}{\textit{Dependent variable:}} \\ 
\cline{2-2} 
\\[-1.8ex] & growth\_rate\_fractional \\ 
\hline \\[-1.8ex] 
 class\_per\_family & 0.007$^{***}$ \\ 
  & (0.0005) \\ 
  & \\ 
 log\_group\_total\_fractional & 0.012$^{***}$ \\ 
  & (0.001) \\ 
  & \\ 
 Constant & $-$0.071$^{***}$ \\ 
  & (0.007) \\ 
  & \\ 
\hline \\[-1.8ex] 
Observations & 752 \\ 
R$^{2}$ & 0.294 \\ 
Adjusted R$^{2}$ & 0.292 \\ 
Residual Std. Error & 0.024 (df = 749) \\ 
F Statistic & 156.150$^{***}$ (df = 2; 749) \\ 
\hline 
\hline \\[-1.8ex] 
\end{tabular} 
\bigskip
 \subcaption*{Section D}
\begin{tabular}{@{\extracolsep{5pt}}lc} 
\\[-1.8ex]\hline 
\hline \\[-1.8ex] 
 & \multicolumn{1}{c}{\textit{Dependent variable:}} \\ 
\cline{2-2} 
\\[-1.8ex] & growth\_rate\_fractional \\ 
\hline \\[-1.8ex] 
 class\_per\_family & 0.014$^{***}$ \\ 
  & (0.001) \\ 
  & \\ 
 log\_group\_total\_fractional & 0.014$^{***}$ \\ 
  & (0.004) \\ 
  & \\ 
 Constant & $-$0.130$^{***}$ \\ 
  & (0.025) \\ 
  & \\ 
\hline \\[-1.8ex] 
Observations & 94 \\ 
R$^{2}$ & 0.498 \\ 
Adjusted R$^{2}$ & 0.487 \\ 
Residual Std. Error & 0.027 (df = 91) \\ 
F Statistic & 45.202$^{***}$ (df = 2; 91) \\ 
\hline 
\hline \\[-1.8ex] 
\end{tabular} 
}
\end{table} 

\begin{table}[!htbp] 
\small
\centering 
  \caption{OLS results with fractional patent counting (B)} 
  \label{B_frac}
  \parbox{.45\linewidth}{
 \subcaption*{Section E}
\begin{tabular}{@{\extracolsep{5pt}}lc} 
\\[-1.8ex]\hline 
\hline \\[-1.8ex] 
 & \multicolumn{1}{c}{\textit{Dependent variable:}} \\ 
\cline{2-2} 
\\[-1.8ex] & growth\_rate\_fractional \\ 
\hline \\[-1.8ex] 
 class\_per\_family & 0.010$^{***}$ \\ 
  & (0.002) \\ 
  & \\ 
 log\_group\_total\_fractional & 0.012$^{***}$ \\ 
  & (0.002) \\ 
  & \\ 
 Constant & $-$0.076$^{***}$ \\ 
  & (0.016) \\ 
  & \\ 
\hline \\[-1.8ex] 
Observations & 114 \\ 
R$^{2}$ & 0.339 \\ 
Adjusted R$^{2}$ & 0.327 \\ 
Residual Std. Error & 0.019 (df = 111) \\ 
F Statistic & 28.428$^{***}$ (df = 2; 111) \\ 
\hline 
\hline \\[-1.8ex] 
\end{tabular} 
\bigskip
  \subcaption*{Section G}
\begin{tabular}{@{\extracolsep{5pt}}lc} 
\\[-1.8ex]\hline 
\hline \\[-1.8ex] 
 & \multicolumn{1}{c}{\textit{Dependent variable:}} \\ 
\cline{2-2} 
\\[-1.8ex] & growth\_rate\_fractional \\ 
\hline \\[-1.8ex] 
 class\_per\_family & 0.019$^{***}$ \\ 
  & (0.003) \\ 
  & \\ 
 log\_group\_total\_fractional & 0.022$^{***}$ \\ 
  & (0.001) \\ 
  & \\ 
 Constant & $-$0.174$^{***}$ \\ 
  & (0.015) \\ 
  & \\ 
\hline \\[-1.8ex] 
Observations & 317 \\ 
R$^{2}$ & 0.444 \\ 
Adjusted R$^{2}$ & 0.440 \\ 
Residual Std. Error & 0.033 (df = 314) \\ 
F Statistic & 125.184$^{***}$ (df = 2; 314) \\ 
\hline 
\hline \\[-1.8ex] 
\end{tabular} 
}
\hfill
  \parbox{.45\linewidth}{
 \subcaption*{Section F}
\begin{tabular}{@{\extracolsep{5pt}}lc} 
\\[-1.8ex]\hline 
\hline \\[-1.8ex] 
 & \multicolumn{1}{c}{\textit{Dependent variable:}} \\ 
\cline{2-2} 
\\[-1.8ex] & growth\_rate\_fractional \\ 
\hline \\[-1.8ex] 
 class\_per\_family & 0.011$^{***}$ \\ 
  & (0.001) \\ 
  & \\ 
 log\_group\_total\_fractional & 0.013$^{***}$ \\ 
  & (0.001) \\ 
  & \\ 
 Constant & $-$0.094$^{***}$ \\ 
  & (0.010) \\ 
  & \\ 
\hline \\[-1.8ex] 
Observations & 478 \\ 
R$^{2}$ & 0.302 \\ 
Adjusted R$^{2}$ & 0.299 \\ 
Residual Std. Error & 0.025 (df = 475) \\ 
F Statistic & 102.613$^{***}$ (df = 2; 475) \\ 
\hline 
\hline \\[-1.8ex] 
\end{tabular} 
\bigskip
  \subcaption*{Section H}
\begin{tabular}{@{\extracolsep{5pt}}lc} 
\\[-1.8ex]\hline 
\hline \\[-1.8ex] 
 & \multicolumn{1}{c}{\textit{Dependent variable:}} \\ 
\cline{2-2} 
\\[-1.8ex] & growth\_rate\_fractional \\ 
\hline \\[-1.8ex] 
 class\_per\_family & 0.018$^{***}$ \\ 
  & (0.002) \\ 
  & \\ 
 log\_group\_total\_fractional & 0.017$^{***}$ \\ 
  & (0.001) \\ 
  & \\ 
 Constant & $-$0.131$^{***}$ \\ 
  & (0.011) \\ 
  & \\ 
\hline \\[-1.8ex] 
Observations & 307 \\ 
R$^{2}$ & 0.470 \\ 
Adjusted R$^{2}$ & 0.466 \\ 
Residual Std. Error & 0.028 (df = 304) \\ 
F Statistic & 134.727$^{***}$ (df = 2; 304) \\ 
\hline 
\hline \\[-1.8ex] 
\end{tabular} 
}
\begin{tabular}{@{\extracolsep{5pt}}lc} 
\hline
\textit{Note:}  & \multicolumn{1}{r}{$^{*}$p$<$0.1; $^{**}$p$<$0.05; $^{***}$p$<$0.01} \\ 
\end{tabular} 
\end{table}

\end{document}